\documentclass[aps, pra, twocolumn, superscriptaddress]{revtex4-2}

\usepackage[pdftex]{graphicx}
\usepackage{amsmath}
\usepackage{amssymb}
\usepackage{physics}
\usepackage{nicefrac}
\usepackage{siunitx}
\usepackage{bbold}
\usepackage{braket}
\usepackage{ragged2e}
\usepackage{lipsum}
\usepackage{multirow}
\setcitestyle{numbers}
\usepackage{layouts}
\usepackage{framed}
\usepackage{float}
\usepackage{bm}
\usepackage{mathtools}
\usepackage{dsfont}
\usepackage{xspace}
\usepackage{microtype}

\usepackage[dvipsnames]{xcolor}
\usepackage{hyperref}
\hypersetup{
  colorlinks   = true, 
  urlcolor     = MidnightBlue, 
  linkcolor    = MidnightBlue, 
  citecolor    = Maroon 
}

\newcommand{\ci}{\text{i}} %complex i
\newcommand{\CZ}{$C\hspace{-1pt}Z$\xspace}
\newcommand{\ZZ}{$Z\hspace{-1pt}Z$\xspace}
\newcommand{\freq}[1]{2\pi\times\SI{#1}{\mega\hertz}}

\begin{document}

\title{Parallel transversal entangling gates for fault-tolerant trapped-ion quantum computing}

\author{Katrin Bolsmann}
\email[Email to ]{k.bolsmann@fz-juelich.de}

\affiliation{Institute for Theoretical Nanoelectronics (PGI-2), Forschungszentrum J\"{u}lich, J\"{u}lich, Germany}
\affiliation{Institute for Quantum Information, RWTH Aachen University, Aachen, Germany}

\author{Alan Kahan}
\affiliation{Instituto de Física Teórica, UAM-CSIC, Universidad Autónoma de Madrid, Cantoblanco, Madrid, Spain }

\author{Alejandro Bermúdez}
\affiliation{Instituto de Física Teórica, UAM-CSIC, Universidad Autónoma de Madrid, Cantoblanco, Madrid, Spain }

\author{Markus Müller}
\affiliation{Institute for Theoretical Nanoelectronics (PGI-2), Forschungszentrum J\"{u}lich, J\"{u}lich, Germany}
\affiliation{Institute for Quantum Information, RWTH Aachen University, Aachen, Germany}

\date{\today}

\begin{abstract}
Fault-tolerant quantum computing requires high-fidelity entangling gates that can be implemented efficiently across multiple qubit pairs. 
We present a scheme for simultaneous pairwise entangling operations, i.e., a transversal controlled-$Z$ gate, across neighboring wells in a double-well trapped-ion architecture. 
A tailored dynamical decoupling scheme suppresses unwanted couplings with a pulse count that scales quadratically with the number of ions.  
We show that incorporating a genuine geometric cancellation reduces the pulse count and, for selected register sizes, allows the parallel operation to be completed faster than an equivalent sequence of two-qubit gates while requiring a significantly smaller two-photon Rabi frequency.
We assess the experimental feasibility of the protocol by analyzing the impact of decoherence and crosstalk arising from laser wave-vector misalignment.
To gain insight into the robustness of fault-tolerant protocols against realistic imperfections of the proposed transversal gate scheme, we investigate how crosstalk-induced gate errors affect the fault-tolerant preparation of a logical Bell state between two $[[4,1,2]]$ quantum-error-detection code blocks.
These results highlight the potential of transversal entangling gates as a building block for trapped-ion-based scalable fault-tolerant quantum error correction.
\end{abstract}

\maketitle

\section{Introduction}
Large-scale quantum computation requires quantum error correction (QEC), where logical qubits are encoded in multiple physical qubits to protect quantum information against errors~\cite{terhal2015, gottesmann2024}.
For this protection to persist during a computation, logical operations must themselves be implemented fault tolerantly, such that a single physical fault cannot propagate into multiple, potentially uncorrectable errors within the same code block.
Transversal gates naturally satisfy this requirement by acting independently on corresponding physical qubits, thereby limiting the propagation of errors within each code block~\cite{eastin2009}.
In particular, a transversal entangling gate between two logical qubits decomposes into independent pairwise entangling operations between corresponding physical qubits of the two code blocks.
Such transversal entangling gates constitute an important primitive in many fault-tolerant (FT) QEC schemes, including constructions based on Calderbank–Shor–Steane (CSS) codes~\cite{steane1997,gottesman1998}.
The potential of performing many such operations in parallel has been particularly evident in neutral-atom processors~\cite{evered2023}, thereby enabling the realization of transversal entangling gates between surface-code blocks with code distances up to $d=7$ in a single parallel step~\cite{bluvstein2024}.
These advances highlight the potential of parallel transversal operations, making their efficient physical implementation an important requirement for scalable FT quantum computation.

Trapped ions provide a well-established platform for quantum information processing, combining high-fidelity operations, long coherence times, and flexible connectivity within individual ion registers~\cite{haeffner2008,bruzewicz2019}.
Experiments with stationary ion registers have demonstrated increasingly advanced FT primitives, including FT logical state preparation and manipulation, universal logical gate operations, repeated rounds of Steane QEC, and more complex QEC protocols based on quantum low-density parity-check (qLDPC) codes~\cite{egan2021,postler2022,postler2024,huang2024, tham2026}.
Scaling trapped-ion systems to larger processors, however, requires architectures that can efficiently connect a growing number of qubits~\cite{kielpinski2002, monroe2013}.
One route uses photonic interconnects to link spatially separated trapped-ion processors and has enabled high-fidelity remote entanglement, distributed quantum computing, and, most recently, distributed QEC~\cite{saha2025,main2025,ainley2026}.
A second approach is based on quantum charge-coupled device (QCCD) architectures, in which ions are shuttled between trapping zones and ion strings are split, merged, and transported through junctions to enable interactions across the processor~\cite{pino2021,ransford2026}.
These architectures have enabled increasingly advanced QEC protocols, including demonstrations of logical performance beyond corresponding physical implementations~\cite{ryananderson2021, ryananderson2024, berthusen2024, hong2024,paetznick2026}, but the required transport and crystal reconfiguration operations add control complexity and execution-time overhead.

\begin{figure*}[t]
    \centering
    \includegraphics[scale=1]{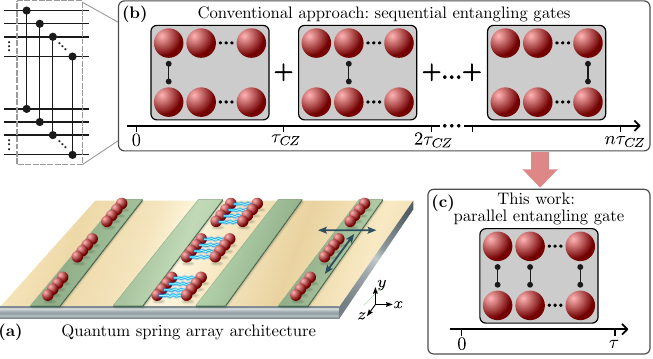}
    \caption{
    Overview of the physical architecture and the parallel implementation of transversal entangling gates considered in this work.
    \textbf{(a)}~Quantum spring array architecture based on a two-dimensional array of spatially separated trapped-ion strings aligned along the axial $z$-direction~\cite{valentini2025}.
    Individual ion strings can be shuttled across the array and moved toward neighboring strings, enabling interactions between the two ion registers through their collective vibrational modes.    %
    \textbf{(b)}~Conventional implementation of a transversal entangling gate between two $n$-qubit registers.
    The $n$ pairwise entangling operations are executed sequentially, resulting in a total duration~$n\tau_{\text{\CZ}}$, provided that the gates are applied consecutively without delays.
    \textbf{(c)}~Parallel implementation proposed in this work.
    All $n$ entangling operations between opposing ions are performed simultaneously within a single gate of duration $\tau$.
}
    \label{fig:intro}
\end{figure*}

An alternative approach is provided by the recently demonstrated quantum spring array architecture~\cite{valentini2025}, consisting of a two-dimensional array of spatially separated ion strings confined in individual trapping wells.
As illustrated in Fig.~\ref{fig:intro}(a), ion strings can be brought into close proximity along both the radial and axial directions, such that  neighboring strings form double wells and interact through their collective vibrational modes without physically merging and splitting them.
A single pairwise entangling operation between two ions in adjacent wells has already been experimentally demonstrated using these inter-well interactions~\cite{wilson2014, valentini2025}.

Realizing a transversal entangling gate between two $n$-qubit registers would require $n$ of these pairwise entangling operations between corresponding ions across the two wells.
A conventional implementation would execute these operations sequentially~\cite{postler2022,ryananderson2022,ryananderson2024,fazio2026}, as illustrated in Fig.~\ref{fig:intro}(b).
Implementing them simultaneously instead offers the prospect of shorter execution times and reduced laser-power requirements, with parallel entangling gates already demonstrated within single ion chains~\cite{figgatt2019}.
Realizing such parallel operations in trapped-ion systems is, however, nontrivial.
Unlike in the neutral-atom implementations, where the spatially rapidly decaying Rydberg interaction naturally provides local selectivity, the collective vibrational modes mediating the inter-well interaction in trapped-ion systems naturally give rise to an all-to-all interaction pattern, making the design of selective multiqubit interactions particularly demanding~\cite{peleg2026}.
A transversal entangling gate specifically requires preserving only the $n$ pairwise interactions between the ions corresponding to the physical qubits of each code block.
The central challenge is therefore to engineer the desired transversal coupling pattern while avoiding additional errors that could compromise the FT character of the transversal operation.

\begin{figure*}[t]
    \centering
    \includegraphics[scale=1]{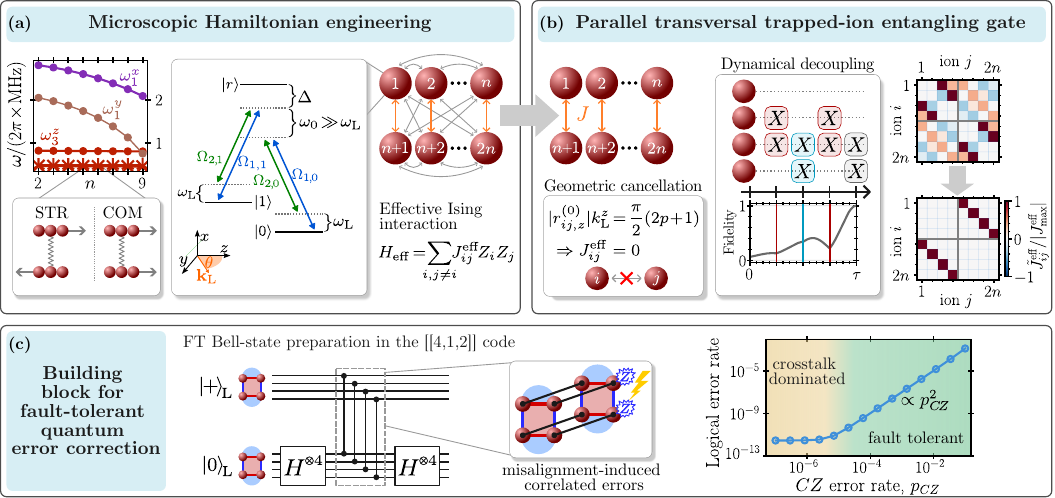}
    \caption{
        Overview of the structure and main results of this work, from microscopic Hamiltonian engineering to the implementation and application of parallel transversal entangling gates in FT quantum computation.
        \textbf{(a)}~Microscopic Hamiltonian engineering in a trapped-ion double-well system.
        The vibrational spectrum of the coupled ion strings as a function of the number of ions per well $n$ shows two closely spaced, lowest-frequency axial modes, the stretch (STR) and center-of-mass (COM) modes, well separated from the next axial mode $\omega_3^z$ (red) and the lowest-frequency $x$ and $y$ modes, $\omega_1^x$ (purple) and $\omega_1^y$ (brown).
        The qubit states $\ket{0}$ and $\ket{1}$, separated by the transition frequency $\omega_0$, are optically coupled to an auxiliary excited state $\ket{r}$ by two far-detuned laser beams with frequencies $\omega_{1,2}$, corresponding Rabi frequencies $\Omega_{1,q}$ and $\Omega_{2,q}$, and single-photon detuning $\Delta$.
        Their interference generates a moving standing wave with beatnote frequency $\omega_\mathrm{L}=\omega_1-\omega_2$ and effective wave vector $\mathbf{k}_\mathrm{L}$, which is chosen to lie in the $y$--$z$ plane and gives rise to the state-dependent force. 
        The COM and STR modes mediate an effective Ising interaction between all $N=2n$ qubits, with strengths $J_{ij}^{\mathrm{eff}}$ determined by the underlying mode structure and spatial configuration of the ions.
        \textbf{(b)}~Engineering of the all-to-all interaction pattern into the pairwise couplings required for the transversal gate.
        Dynamical decoupling (DD) uses local $X$-pulses to reverse selected interactions such that they cancel over the gate evolution.
        The shown pulse sequence illustrates the DD scheme for $n=4$ ions per well, resulting in a gate fidelity that approaches unity at the end of the ideal evolution.
        Geometric cancellation (GC) can complement the DD scheme by suppressing selected couplings directly at the Hamiltonian level through the spatial dependence of the interaction, thereby reducing the required pulse count and gate duration.
        The heatmaps at the right show the interaction matrices before and after applying the interaction-engineering scheme, where $\widetilde{J}_{ij}^{\mathrm{eff}}$ (cf. Eq.~\eqref{eqn:acc_interaction}) denotes the effective coupling obtained by averaging $J_{ij}^{\mathrm{eff}}$ over the sign changes induced by the DD sequence.
        After interaction engineering, only the couplings between opposing ions required for the transversal gate remain, while the unwanted interactions are suppressed.
        \textbf{(c)}~Application of the transversal entangling gate as an exemplary resource for FT quantum computation, considering the preparation of a logical Bell state between two instances of the $[[4,1,2]]$ code.
        Wave-vector misalignment prevents complete GC and gives rise to residual correlated $ZZ$ crosstalk errors.
        Intra-block errors can propagate into an undetectable logical $X$ error, causing the characteristic quadratic scaling with the physical \CZ error rate $p_{\text{\CZ}}$ to cross over to a crosstalk-dominated error floor at small $p_{\text{\CZ}}$.
    }
    \label{fig:overview}
\end{figure*}

Building on the parallel transversal-gate concept discussed in Ref.~\cite{valentini2025}, we theoretically propose a scheme for implementing a parallel transversal entangling gate between two ion registers of, in principle, arbitrary size confined in radially separated trapping wells and characterize its resource requirements and sensitivity to experimentally relevant error sources.
An overview of the paper structure and the main results is shown in Fig.~\ref{fig:overview}.
We first characterize the motional mode structure of the double-well system in Sec.~\ref{sec:radial_trapping}, identifying the spectrally isolated axial center-of-mass (COM) and stretch (STR) modes as the relevant collective modes mediating the inter-well interaction.
In Sec.~\ref{sec:Hamiltonian}, we introduce the internal qubit structure and laser configuration illustrated in Fig.~\ref{fig:overview}(a), where a pair of off-resonant laser beams generates a state-dependent force coupling the internal qubit states to the collective COM and STR motion of the ions.
From this microscopic description, we derive an effective Ising Hamiltonian governing the phonon-mediated qubit--qubit interactions.

To obtain the coupling pattern required for a transversal entangling gate, Sec.~\ref{sec:Interaction_Engineering} introduces two complementary techniques: dynamical decoupling (DD) and geometric cancellation (GC), both shown schematically in Fig.~\ref{fig:overview}(b).
The DD scheme selectively suppresses all unwanted interactions while preserving the couplings between opposing ions, requiring $\mathcal{O}(n)$ free-evolution intervals under the all-to-all coupling and $\mathcal{O}(n^2)$ single-qubit $\pi$-pulses.
Additionally, GC exploits the spatial dependence of the effective interaction to eliminate a subset of unwanted couplings directly at the Hamiltonian level, thereby reducing both the number of required DD pulses and the resulting gate duration.
We first characterize the resulting transversal gate in Sec.~\ref{sec:results} in the absence of additional error sources, analyzing the intrinsic infidelity arising from the nonuniform mode structure of the radially coupled inhomogeneous ion chains together with the gate duration and Rabi-frequency requirements.
For many of the qubit register sizes considered here, combining DD and GC enables faster transversal gates than sequential single-cycle pair gates, while requiring a smaller effective two-photon Rabi frequency.

Finally, Sec.~\ref{sec:error_analysis} investigates the robustness of the gate against experimentally relevant imperfections, including entangling pulse-area fluctuations, residual coupling to spectator vibrational modes, qubit dephasing, and crosstalk arising from laser wave-vector misalignment.
In particular, imperfect GC due to wave-vector misalignment generates correlated intra- and inter-well \ZZ interactions that extend beyond the intended transversal coupling pattern and is especially relevant in an FT setting.
To assess the relevance of these crosstalk errors for FT quantum computation, we consider the preparation of a logical Bell state between two $[[4,1,2]]$ error-detecting code blocks, as illustrated in Fig.~\ref{fig:overview}(c).
While most individual crosstalk events remain detectable or act trivially on the encoded state, a subset of correlated intra-well errors can induce undetectable logical errors and give rise to a crosstalk-dependent floor in the logical error rate.

Overall, we propose a scheme for implementing parallel transversal entangling gates in a trapped-ion double-well architecture by efficiently engineering the underlying interaction pattern.
Our results show that this approach can reduce both gate duration and required Rabi frequency compared with sequential implementations.
They also highlight the importance of controlling implementation-specific error sources, such as qubit dephasing and wave-vector misalignment, which can limit the gate performance and potentially compromise its FT properties.
Taken together, these results establish parallel transversal entangling gates as a promising resource for FT quantum information processing with trapped ions.

\section{Parallel transversal gate implementation}
In this section, we introduce a protocol for realizing a parallel transversal entangling gate between two ion strings confined in radially separated trapping wells. 
The scheme is inspired by earlier proposals exploiting anisotropic spin-spin interactions in trapped-ion systems to engineer tailored coupling patterns \cite{bermudez2011, bermudez2012}. 
We begin by introducing the double-well architecture and deriving the effective Ising Hamiltonian governing the qubit interactions.
Building on this, we construct a dynamical decoupling (DD) sequence that selectively cancels unwanted interactions while preserving the couplings between opposing ions in the two wells, with a pulse count that scales quadratically with the number of ions.
Combined with geometric interaction engineering, which exploits the directional dependence of the interaction to suppress a subset of couplings already at the Hamiltonian level, the complexity and pulse count of the DD sequence can be further reduced.
Finally, we show how the resulting interaction pattern naturally implements a transversal entangling gate between the two ion strings and discuss the intrinsic fidelity limitations arising from the mode structure in the system.

\subsection{Radially separated ion chains}\label{sec:radial_trapping}
Following Ref.~\cite{valentini2025}, we consider two trapped-ion strings aligned along the axial $z$-direction and confined in a symmetric double-well potential, with the wells radially separated along the $x$-axis,
\begin{equation}
\Phi_\mathrm{T}(\mathbf{R})=\frac{1}{2}\alpha R_{x}^2+\frac{1}{4!}\beta R_{x}^4 + \frac{1}{2}\phi_y R_{y}^2 + \frac{1}{2}\phi_z R_{z}^2,
\end{equation}
where $\alpha<0$ and $\beta>0$ are the second- and fourth-order curvatures along the well axis and $\phi_{y,z}$ are the second-order curvatures along the $y$- and $z$-axes, respectively.
The distance $d=\sqrt{-24\alpha/\beta}$ between the two wells is determined by the location of the minima along the $x$-axis.
We further define the anisotropy parameters $\kappa_a = (\omega_z/\omega_a)^2$ with $a\in \{x, y\}$ and the frequencies $\omega_x^2 = -e\alpha/M,\, \omega_y^2 = e\phi_y/M,$ and $\omega_z^2 = e\phi_z/M$ with the ion mass $M$.
The total potential energy of $N$ ions in the trap is composed of the trap potential and the Coulomb repulsion between the ions,
\begin{equation}
    \begin{split}
        V = &\sum_{i=1}^N \Big(\frac{\ell^2}{\kappa_x d^2}r^4_{i,x} - \frac{1}{2\kappa_x}r^2_{i,x}  + \frac{1}{2\kappa_y}r^2_{i,y} + \frac{1}{2}r^2_{i,z}\Big ) \\
        &+ \frac{1}{2}\sum_{i=1}^N\sum_{j\neq i}\frac{1}{|\mathbf{r}_i - \mathbf{r}_j|},
    \end{split}
\end{equation}
where we use dimensionless units $\mathbf{r}_i = \mathbf{R}_i/\ell$ with the characteristic length scale $\ell^3 = {e^2}/(4\pi \epsilon_0 M \omega_z^2)$ and measure energies in units of $M\omega_z^2\ell^2$. 
The equilibrium positions $ \mathbf{r}_i^{(0)}$ are defined by the potential minimum, i.e., they satisfy $\nabla_{\mathbf{r}_i}V|_{\mathbf{r}_i^{(0)}} = 0$ for all $i=1, 2, ..., N$.
For sufficiently strong radial confinement and moderate ion numbers, the resulting equilibrium configuration consists of two opposing mirror-symmetric ion strings with nonuniform spacing, where the ions are closer together at the center than at the edges.

We consider small displacements around the equilibrium positions, $\mathbf{r}_i(t)=\mathbf{r}_i^{(0)}+\delta\mathbf{r}_i(t)$, and expand the potential to second order,
\begin{equation}
V\approx\frac{1}{2}\sum_{i,j}\sum_{a,b}H_{ij}^{ab}\,\delta r_{i,a}\delta r_{j,b},\;\; H_{ij}^{ab}\!=\!\left.\frac{\partial^2V}{\partial r_{i,a}\partial r_{j,b}}\right|_{\mathbf{r}^{(0)}}\!.
\end{equation}
Here, \(i,j\in\{1,\ldots,N\}\) label the ions, \(a,b\in\{x,y,z\}\) denote the spatial directions, and \(H_{ij}^{ab}\) is the corresponding element of the Hessian matrix evaluated at the equilibrium configuration.
The trap potential is separable in the three spatial directions and therefore contributes only to Hessian elements with \(a=b\). 
In contrast, the Coulomb interaction is generally not separable and gives rise to mixed second derivatives that couple different spatial directions in the Hessian.
For $a\neq b$, these coupling terms are proportional to
\begin{equation}
H_{ij}^{ab}\propto\frac{r_{ij,a}^{(0)}r_{ij,b}^{(0)}}{|\mathbf r_{ij}^{(0)}|^5}
\end{equation}
with $\mathbf r_{ij}^{(0)}=\mathbf r_j^{(0)}-\mathbf r_i^{(0)}$, and do not vanish in general.
For the planar equilibrium configurations considered here, $r_{i,y}^{(0)}\approx 0$ for all ions. 
Consequently, all Hessian elements coupling the $y$-direction to the $x$- or $z$-directions are negligibly small.
Arranging the displacements of all $N$ ions in the vector $\delta\mathbf{r}=(\delta r_{1,x},\ldots,\delta r_{N,x},\delta r_{1,y},\ldots,\delta r_{N,y},\delta r_{1,z},\ldots,\delta r_{N,z})^{\mathsf T}$, the Hessian takes the block form
\begin{equation}
H=
\begin{pmatrix}
H^{xx} & 0 & H^{xz}\\
0 & H^{yy} & 0\\
(H^{xz})^{\mathsf T} & 0 & H^{zz}
\end{pmatrix}.
\end{equation}
The remaining mixed block $H^{xz}$ is governed by products $r_{ij,x}^{(0)}r_{ij,z}^{(0)}$. 
For ion pairs within the same string, ${r_{ij,x}^{(0)}\approx0}$, whereas for vertically aligned ions in opposite wells, $r_{ij,z}^{(0)}\approx0$. 
Hence, only diagonally displaced inter-well ion pairs contribute to $H^{xz}$, and these contributions are additionally suppressed by the large inter-well distance through the factor $|\mathbf r_{ij}^{(0)}|^{-5}$. 
The mixed block is therefore much smaller than the diagonal blocks $H^{xx}$ and $H^{zz}$.
In our system, we assume strong radial confinement $\kappa_x \approx \kappa_y\ll1$ to ensure the formation of two parallel linear ion strings. 
The resulting separation between the radial and axial mode frequencies, together with the weak residual $xz$-coupling, leads to only negligible hybridization of the normal modes.
Accordingly, we neglect the remaining off-diagonal block $H^{xz}$ and treat the $x$- and $z$-directions independently. Defining
\begin{equation}\label{eqn:Hesse-matrix}
H_{ij}^{a}=\left.\frac{\partial^2V}{\partial r_{i,a}\partial r_{j,a}}\right|_{\mathbf r^{(0)}}, \qquad a\in\{x,y,z\},
\end{equation}
the normal modes are obtained by solving the eigenequation
\begin{equation}
H^{a}\mathcal{M}_m^{a}=\lambda_m^{a}\mathcal{M}_m^{a}.
\end{equation}
The corresponding frequency of the normal mode with eigenvector $\mathcal{M}_m^{a}$ is then given by $\omega_m^{a}=\omega_z\sqrt{\lambda_m^{a}}$.

In the following, we set $\hbar=1$ and quantize the vibrational degrees of freedom by introducing the bosonic annihilation and creation operators $b_m^a$ and $(b_m^a)^\dagger$ for the $m$th normal mode in $a$-direction, which satisfy the canonical commutation relations $[b_m^a,(b_n^b)^\dagger]=\delta_{mn}\delta_{ab}$.
The corresponding canonical transformation
\begin{equation}\label{eqn:can_trafo}
    \delta r_{i,a} = \sum_{m=1}^N \frac{\mathcal{M}_{i,m}^{a}}{\sqrt{2M\omega_m^a}\,\ell}(b_m^a + (b_m^a)^\dagger)
\end{equation}
allows us to express the phononic Hamiltonian of the vibrational degrees of freedom as
\begin{equation}\label{eqn:phonon_hamiltonian}
    H_\mathrm{p} = \sum_{m=1}^N\sum_{a}\omega_m^a\left((b_m^a)^\dagger b_m^a + \frac{1}{2}\right).
\end{equation}

The two lowest-energy modes correspond to axial center-of-mass motions of the individual ion strings. 
Among them, the lower-frequency mode is the stretch~(STR) mode, in which the two strings oscillate out of phase, while the center-of-mass~(COM) mode, with $\omega_\mathrm{COM}^z=\omega_z$,  corresponds to an in-phase oscillation. 
While the splitting $\Delta\omega_\mathrm{c} = \omega_\mathrm{COM}^z-\omega_\mathrm{STR}^z$ between these two modes is small, they are well separated from all remaining axial and radial modes, as illustrated in Fig.~\ref{fig:overview}(a).
The spectral isolation of the STR and COM modes allows them to be addressed selectively, making them ideally suited for mediating the interaction between the qubits underlying the transversal gate, as discussed in the following section.

\subsection{Derivation of the effective interaction Hamiltonian}\label{sec:Hamiltonian}
In the following, we derive the effective Ising Hamiltonian governing the qubit interactions, closely following the approach of Ref.~\cite{bermudez2011}. 
We adapt the derivation, originally developed in the context of frustrated spin-ladder models in planar trapped-ion crystals, to the double-well geometry and vibrational mode structure considered here.
We consider two internal atomic states $\ket{0_i}$ and $\ket{1_i}$ defining the qubit subspace of the $i$th ion and an auxiliary excited state~$\ket{r_i}$ which can be optically coupled to both qubit states.
To engineer an Ising-type interaction, a pair of laser beams with optical frequencies $\omega_1$ and $\omega_2$ is applied to the ions, as illustrated in Fig.~\ref{fig:overview}(a).
Their interference generates a moving standing wave with beatnote frequency $\omega_\mathrm{L}=\omega_1-\omega_2$ and effective dimensionless wave vector $\mathbf{k}_\mathrm{L}=\mathbf{k}_1-\mathbf{k}_2$, where all wave vectors are expressed in units of $\ell^{-1}$.
We consider a regime where the lasers are far off-resonant from the transitions between the qubit subspace and the excited state.
Specifically, the detuning~$\Delta$ is assumed to be much larger than the excited-state decay rate $\Gamma_r$ and the laser Rabi frequencies $\Omega_{1,q}$ and $\Omega_{2,q}$, with $q\in\{0,1\}$, such that the excited state can be adiabatically eliminated.
Further assuming that the beatnote is far detuned from the qubit transition, $\omega_\mathrm{L}\ll\omega_0$, and chosen close to the axial motional frequencies, $\omega_\mathrm{L}\approx\omega_z$, eliminating $\ket{r_i}$ results in an effective two-level system with a time-dependent laser-induced AC Stark shift described by the dipole Hamiltonian 
\begin{equation}\label{eqn:dipole_hamiltonian}
    H_\mathrm{d} = \frac{\Omega_\mathrm{L}}{2}\sum_i e^{\ci\mathbf{k}_\mathrm{L}\cdot\mathbf{r}_i^{(0)}}e^{\ci(\mathbf{k}_\mathrm{L}\cdot\delta\mathbf{r}_i - \omega_\mathrm{L}t)}Z_i + \mathrm{H.c.},
\end{equation}
where $Z_i = \ket{0_i}\!\!\bra{0_i}-\ket{1_i}\!\!\bra{1_i}$ is the Pauli-$Z$ operator acting on the $i$th qubit and $\Omega_\mathrm{L}=(\Omega_{1,1}\Omega_{2,1}^* - \Omega_{1,0}\Omega_{2,0}^*)/(4\Delta)$ is the two-photon differential Rabi frequency.
Without loss of generality, we choose the laser phases such that $\Omega_\mathrm{L}$ is real.
The adiabatic elimination also gives rise to effective dissipative processes associated with off-resonant Rayleigh and Raman photon scattering through the virtually populated excited state.
For the large detunings considered here, these dissipative processes are strongly suppressed and are neglected in the following.
In addition, the elimination produces static AC Stark shifts and a state-independent running-wave contribution that couples to the vibrational modes.
Both contributions are assumed to be carefully compensated experimentally by an appropriate choice of the laser parameters.
Further details and a comprehensive derivation of Hamiltonian~\eqref{eqn:dipole_hamiltonian} based on the effective operator formalism for open quantum systems introduced in Ref.~\cite{reiter2012} can be found in App.~\ref{app:dipole-Hamiltonian-derivation}.

We choose the laser wave vector to lie in the $y$-$z$-plane, $\mathbf{k}_\mathrm{L} = k\sin\theta\mathbf{e}_y + k \cos\theta\mathbf{e}_z$, as illustrated in Fig.~\ref{fig:overview}(a), such that the laser couples only to the axial $z$- and radial $y$-modes.
After applying the canonical transformation~\eqref{eqn:can_trafo} and going into the interaction picture with respect to the phonon Hamiltonian $H_\mathrm{p}$~\eqref{eqn:phonon_hamiltonian}, we expand the dipole Hamiltonian~\eqref{eqn:dipole_hamiltonian} for small Lamb-Dicke parameters ${\eta_m^a = k_\mathrm{L}^a/(\!\sqrt{2M\omega_m^a}\,\ell)\ll1}$ with $a\in\{y,z\}$ and get
\begin{equation}
\begin{aligned}
H_\mathrm{d}\! &= \frac{\Omega_\mathrm{L}}{2}\!\sum_i e^{\ci\mathbf{k}_\mathrm{L}\cdot\mathbf{r}_i^{(0)}}e^{-\ci\omega_\mathrm{L}t}\\
&\,\,\times\! \bigg\{\!1\! + \ci\!\!\sum_{m=1}^{N}\!\! \Big[ \eta_m^{y}\mathcal{M}^{y}_{i,m}\!\big(e^{-\ci\omega_m^{y}t} b_m^y\!+ e^{\ci\omega_m^{y}t} (b_m^y)^\dagger\big)\\
&\,\,+ \eta_m^{z}\mathcal{M}^{z}_{i,m}\!\big(e^{-\ci\omega_m^{z}t} b_m^z\!+ e^{\ci\omega_m^{z}t} (b_m^z)^\dagger\big)\! \Big]\! \!\bigg\} Z_i + \mathrm{H.c.}
\end{aligned}
\end{equation}
We choose the laser beatnote close to the two spectrally isolated axial modes, $\omega_\mathrm{L}\approx\omega_\mathrm{COM}^z,\omega_\mathrm{STR}^z$, and define the mode detunings $\delta_m^a=\omega_m^a-\omega_\mathrm{L}$.
Since the COM and STR modes are well separated from the remaining axial and radial branches, their detunings are much smaller than those of the remaining modes,
\begin{equation}
    |\delta_\mathrm{COM}^z|,\,|\delta_\mathrm{STR}^z|\ll|\delta_m^z|,\,|\delta_m^y|,
\end{equation}
where $m$ labels spectator modes only.
In addition, we require
\begin{equation}\label{eqn:RWAconditions}
    \Big|\frac{\Omega_\mathrm{L}}{\omega_\mathrm{L}}\Big|\ll1, \quad \Big|\frac{\eta_m^a\Omega_\mathrm{L}}{\delta_m^a}\Big| \ll \Big|\frac{\eta_\mathrm{STR}^z\Omega_\mathrm{L}}{\delta_\mathrm{STR}^z}\Big|,\Big |\frac{\eta_\mathrm{COM}^z\Omega_\mathrm{L}}{\delta_\mathrm{COM}^z}\Big|.
\end{equation}
The first condition allows us to neglect the rapidly oscillating contributions, while the second ensures that the off-resonant coupling to the spectator modes is small compared with the coupling to the targeted COM and STR modes.
We apply a rotating-wave approximation and neglect the residual coupling to the spectator modes.
The dipole Hamiltonian then reduces to the interaction mediated by the axial COM and STR modes
\begin{equation}\label{eqn:dipole_Hamiltonian_LDA}
    H_\mathrm{d} = \sum_{i,m} g_{i,m} e^{\ci\delta_m^{z} t} (b_m^z)^\dagger Z_i + \mathrm{H.c.}
\end{equation}
with $g_{i,m} = \frac{\ci}{2}\Omega_\mathrm{L}\eta_m^z e^{\ci\mathbf{k}_\mathrm{L}\cdot\mathbf{r}_i^{(0)}}\mathcal{M}_{i,m}^z$ and $m\in\{\mathrm{COM},\mathrm{STR}\}$.
From this point onward, we restrict the vibrational dynamics to these two axial modes and omit the superscript $z$ for notational simplicity.
Accordingly, the phonon Hamiltonian in Eq.~\eqref{eqn:phonon_hamiltonian} and all subsequent mode sums are understood to be restricted to the resulting two-mode subspace unless stated otherwise.
The effect of the finite off-resonant coupling to the neglected spectator modes is analyzed separately in Sec.~\ref{sec:spectator_mode_coupling}.

The dipole Hamiltonian in Eq.~\eqref{eqn:dipole_Hamiltonian_LDA} gives rise to an effective Ising-type interaction between the qubits mediated by the exchange of virtual phonons of the considered axial modes.
In particular, the state-dependent force couples the internal qubit states to the collective vibrational modes, such that phonon excitations generated by one ion can be reabsorbed by another and thereby mediate an effective qubit--qubit interaction.
The resulting dynamics can be obtained exactly by performing a Magnus expansion of the dipole Hamiltonian, as detailed in App.~\ref{app:magnus-expansion}.
The corresponding propagator contains a state-dependent displacement of the motional modes as well as a geometric phase giving rise to the effective qubit--qubit interaction.
At times when the phase-space trajectories of the driven STR and COM modes are closed, the displacement vanishes and the qubit and vibrational degrees of freedom disentangle, as discussed in more detail in Sec.~\ref{sec:gate_parameters}.
The remaining evolution is then governed by the effective interaction Hamiltonian
\begin{equation}
\begin{gathered}\label{eqn:effective_hamiltonian}
H_\mathrm{eff} =  \sum_{i=1}^N \sum_{j\neq i} J_{ij}^{\mathrm{eff}}Z_i Z_j,
\quad
J_{ij}^{\mathrm{eff}}=\chi_{ij}\cos\bigl(\mathbf{k}_{\mathrm{L}}\!\cdot\mathbf{r}_{ij}^{(0)}\bigr),
\\
\chi_{ij} = -\sum_m\frac{\Omega_{\mathrm{L}}^{2}\eta_m^2}
{4\delta_m}
\mathcal{M}_{i,m}\mathcal{M}_{j,m}.
\end{gathered}
\end{equation}
This effective Hamiltonian describes an Ising-type interaction which is mediated by the collective vibrational STR and COM modes of the two coupled ion strings. 
The coupling strengths are determined jointly by the mode-dependent prefactor $\chi_{ij}$ and the geometric phase factor~$\cos\bigl(\mathbf{k}_{\mathrm{L}}\!\cdot\mathbf{r}_{ij}^{(0)}\bigr)$, which together set both the magnitude and sign of the effective couplings between different ion pairs.
In the following section, we introduce a scheme to engineer this interaction pattern such that the desired couplings required to implement a transversal entangling gate are retained while all unwanted interactions are suppressed.

\subsection{Interaction engineering}\label{sec:Interaction_Engineering}
The effective Hamiltonian in Eq.~\eqref{eqn:effective_hamiltonian} generally gives rise to interactions across the full ion register, where the individual coupling strengths are set by the normal-mode structure and the spatial phase dependence entering $J_{ij}^{\mathrm{eff}}$.
For the transversal gate considered here, however, we require $n=N/2$ independent, equally strong couplings, $J_{i,i+n}^\mathrm{eff}=J$, between opposing ions $i$ and $i+n$ in the two wells, while all other couplings vanish, as illustrated in Fig.~\ref{fig:overview}(b).
In this case, the Hamiltonian decomposes into $n$ non-interacting subspaces described by $H_i^\mathrm{sub}=2J Z_iZ_{i+n}$, where the factor of two results from the sums in Eq.~\eqref{eqn:effective_hamiltonian} counting each ion pair twice.
The time evolution of the total system is then described by 
\begin{equation}\label{eqn:time_evolution}
    U(t)=\prod_{i=1}^n \exp(-2\ci J Z_i Z_{i+n} t),
\end{equation}
where, up to single-qubit $Z$-rotations, $U(\tau)$ implements a maximally entangling \CZ gate between each pair of opposing ions from each individual well after an evolution time $\tau = \pi/(8|J|)$.

The central task is therefore to realize this coupling pattern starting from the generally nonuniform interaction generated by $H_\mathrm{eff}$. 
In the following, we first construct a DD scheme that suppresses all unwanted couplings while preserving the interactions between opposing ion pairs. 
Having established this scheme, we then show how the geometric dependence of $J_{ij}^{\mathrm{eff}}$ can be exploited to cancel a subset of the unwanted interactions already at the Hamiltonian level, thereby reducing the number of decoupling pulses required and making the overall interaction engineering more efficient.

\begin{figure*}[t]
            \centering
            \includegraphics[scale=1]{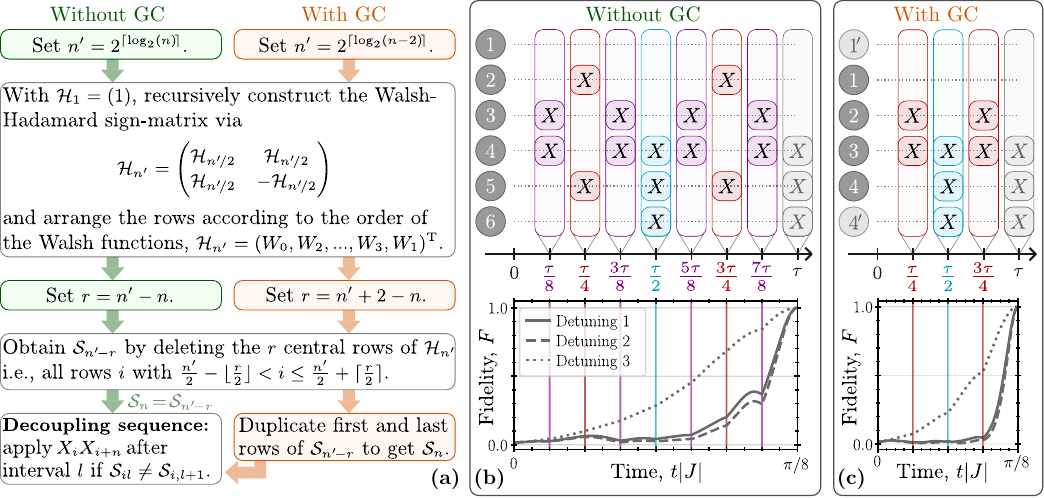}
\caption{
    Construction and application of the interaction-engineering schemes.
    \textbf{(a)}~Step-by-step construction of the sign matrix $\mathcal{S}_n$ from Walsh--Hadamard matrices for dynamical decoupling (DD) alone and DD combined with geometric cancellation (GC) for $n\geq4$, highlighted in green and orange, respectively.
    The decoupling sequence applies $X_iX_{i+n}$ to the $i$th qubit and its opposing qubit $i+n$ at the $l$th time step whenever the corresponding row of the sign matrix changes sign from $l$ to $l+1$.
    \textbf{(b)}~Pulse sequence obtained from $\mathcal{S}_6$ in Eq.~\eqref{eqn:S6} for $n=6$ ions per well using DD alone.
    Only one well is shown, since identical pulses are applied simultaneously to the opposing ions in the other well.
    The final gray pulses restore the initial qubit basis frame but are not required for the decoupling itself.
    The lower panel shows the corresponding state fidelity of the transversal \CZ gate for the three detuning scenarios (cf.~Fig.~\ref{fig:results_fidelity}(a) and Sec.~\ref{sec:gate_parameters}), all reaching a final infidelity below $10^{-3}$ at the gate time $\tau=\pi/(8|J|)$ in the absence of additional error sources.
    \textbf{(c)}~Analogous pulse sequence and gate fidelity for $n=6$ when DD is combined with GC.
    The sequence is based on $\mathcal{S}_4$, with the pulse sequences of the first and last ions duplicated for the additional ions $1'$ and $4'$, while their residual interactions are suppressed by GC.
    This reduces the number of time intervals from eight to four and the pulse count per well from $15$ to $7$, while all three detuning scenarios retain a final infidelity below $10^{-3}$.
    }
    \label{fig:decoupling_scheme}
\end{figure*}

\subsubsection{Dynamical decoupling}
We first introduce a general scheme for a complete decoupling of an arbitrary number of qubits by applying a specific sequence of $X$-pulses. 
The scheme requires a number of time intervals that grows linearly with the number of qubits, while the total number of applied pulses scales quadratically. 
As we show below, this scheme can be straightforwardly adapted to our double-well system, where instead of suppressing all interactions, we selectively retain the interactions between opposing ions while canceling all unwanted couplings.

The general construction is based on a well-established approach to decoupling pairwise interactions using Walsh--Hadamard matrices~\cite{jones1999, leung2000, hayes2011, bhole2020, votto2024}. 
These square matrices provide a particularly natural decoupling scheme for systems with $n=2^m$ particles using the same number of equal time intervals.
First, we note that an $X$-pulse applied to a qubit evolving under an Ising-type Hamiltonian reverses the sign of all interactions involving this qubit, since $X_i Z_i X_i=-Z_i$. 
If we divide the total gate time into an even number of equally spaced time intervals, complete decoupling can be achieved by choosing the pulse sequence such that every pairwise interaction has a positive sign for half of the total evolution time and a negative sign for the other half, giving rise to the $n=2^m$ constraint in the Walsh--Hadamard construction. 
In this case, the accumulated interaction averages to zero at the end of the gate.

We represent the pulse sequence by an $n\times k$ sign matrix $\mathcal{S}_n$, where each row corresponds to a qubit and each column to one of the $k$ time intervals.
The entries $\mathcal{S}_{il}\in\{+1,-1\}$ indicate the sign of $Z_i$ during time interval $l$. 
Whenever two entries have the same sign during a given interval, i.e., $\mathcal{S}_{il} = \mathcal{S}_{jl}$, the interaction between qubits $i$ and $j$ retains its original sign, whereas $\mathcal{S}_{il} \neq \mathcal{S}_{jl}$ reverses the interaction. 
Hence, during interval $l$, the interaction $J_{ij}^\mathrm{eff}Z_iZ_j$ acquires the sign~$\mathcal{S}_{il}\mathcal{S}_{jl}$. 
The effective interaction strength
\begin{equation}\label{eqn:acc_interaction}
    \widetilde{J}_{ij}^{\mathrm{eff}}
    =
    \frac{J_{ij}^{\mathrm{eff}}}{k}
    \sum_{l=1}^{k}
    \mathcal{S}_{il}\mathcal{S}_{jl}
\end{equation}
between qubits $i$ and $j$ therefore vanishes whenever $\sum_{l=1}^{k}\mathcal{S}_{il}\mathcal{S}_{jl}=0$. 
Accordingly, all $n$ qubits are decoupled if all rows of $\mathcal{S}_n$ are pairwise orthogonal, i.e., $\mathcal{S}_n\mathcal{S}_n^{\mathrm{T}}= k\mathds{I}$.

A family of matrices satisfying this condition is given by the Walsh--Hadamard matrices~\cite{horadam2007, bhole2020, votto2024}. 
In general, a Hadamard matrix $\mathcal{H}_n$ is an $n\times n$ matrix with entries $\mathcal{H}_{il}\in\{+1,-1\}$ whose rows and columns are mutually orthogonal, such that $\mathcal{H}_n\mathcal{H}_n^{\mathrm{T}}=n\mathds{I}$~\cite{horadam2007}. 
This is precisely the orthogonality condition required for complete decoupling in the sign-matrix representation introduced above. 
Walsh--Hadamard matrices form a particular family of Hadamard matrices whose rows can be identified with discrete Walsh functions~\cite{walsh1923, beer1981}. 
Starting from $\mathcal{H}_1=(1)$, the Walsh--Hadamard matrix for every dimension $n=2^m$ can be constructed recursively according to~\cite{sylvester1867, leung2000, horadam2007}
\begin{equation}\label{eqn:H_construction}
    \mathcal{H}_{2n}=\begin{pmatrix}\mathcal{H}_n&\mathcal{H}_n\\\mathcal{H}_n&-\mathcal{H}_n\end{pmatrix}. 
\end{equation}
Therefore, whenever the number of qubits equals a power of two, the corresponding Walsh--Hadamard matrix can be directly used as the sign matrix of the decoupling sequence, $\mathcal{S}_n=\mathcal{H}_n$.
For instance,
\begin{equation}\label{eqn:S4}
   \mathcal{S}_4 = \mathcal{H}_4=\begin{pmatrix}
        +1&+1&+1&+1\\
        +1&-1&+1&-1\\
        +1&+1&-1&-1\\
        +1&-1&-1&+1
    \end{pmatrix}
    =\begin{pmatrix}W_0\\W_3\\W_1\\W_2\end{pmatrix},
\end{equation}
where we label the Walsh functions $W_p$ such that the index $p$ corresponds to the number of sign changes in the respective row and therefore to the number of $X$-pulses applied to the corresponding qubit during the sequence, excluding a possible final pulse required to restore the original qubit basis frame.

We can directly translate the sign matrix~$\mathcal{S}_4$ into a decoupling sequence for four qubits consisting of four equal time intervals.
After interval $l$, an $X$-pulse is applied to ion~$i$ whenever the corresponding row changes sign between the two columns associated with two consecutive time intervals, i.e., $\mathcal{S}_{il}\neq\mathcal{S}_{i,l+1}$. 
Including the final pulses required to remove the net bit flips, the resulting evolution can be written as
\begin{equation}
    X_2X_3U\Big(\frac{\tau}{4}\Big)X_2X_4U\Big(\frac{\tau}{4}\Big)X_2X_3U\Big(\frac{\tau}{4}\Big)X_2X_4U\Big(\frac{\tau}{4}\Big).
\end{equation}
Setting $\tau=0$ makes explicit that the final $X_2X_3$ pulses ensure that the complete pulse sequence reduces to the identity, leaving no net qubit flips after the decoupling sequence.

The construction of $\mathcal{S}_n$ can be extended to an arbitrary number of qubits $n$ by first generating the decoupling scheme for the next power of two, $n'=2^{\lceil\log_2(n)\rceil}$, using Eq.~\eqref{eqn:H_construction} and subsequently removing $r=n'-n$ rows from the corresponding sign matrix.
Since removing rows does not affect the mutual orthogonality of the remaining rows, the resulting sign matrix $\mathcal{S}_n$ still provides complete decoupling of the remaining $n$ qubits. 
To minimize the total number of pulses, it is most efficient to remove the rows with the largest number of sign changes, corresponding to the Walsh functions with the highest indices $p$. 
For example, to construct a decoupling scheme for three qubits, we start from $\mathcal{S}_4$ in Eq.~\eqref{eqn:S4} and remove the second row, which corresponds to $W_3$ and therefore requires the largest number of pulses.
Hence, the total number of pulses required to decouple $n$ qubits is given by
\begin{equation}
    N_X = \sum_{p=0}^{n-1} p = \frac{n(n-1)}{2} = \mathcal{O}(n^2),
\end{equation}
while the number of required time intervals is $n' = \mathcal{O}(n)$.
Here, we exclude the $\lfloor n/2 \rfloor$ additional $X$-pulses needed to restore the initial qubit basis frame, as they do not affect the overall scaling of the decoupling scheme. 
These final corrections must be accounted for, but may either be applied explicitly or incorporated in post-processing, depending on the specific application. 
We therefore omit them whenever referring to the pulse count in the following.

So far, we have discussed well-established refocusing techniques for fully decoupling qubits evolving under an effective Ising Hamiltonian with arbitrary long-range couplings.
We now exploit the underlying principles to construct an efficient decoupling scheme for the double-well system containing a total of $N$ ions, which selectively preserves the transversal couplings between opposing ions while suppressing all other interactions, thereby realizing the interaction pattern required for an FT transversal entangling gate.
To this end, we construct the decoupling scheme introduced above for a single ion well containing $n=N/2$ ions and apply each $X$-pulse simultaneously to the corresponding pair of opposing ions, i.e., $X_iX_{i+n}$. 
In terms of the sign matrix, this amounts to assigning the same row to each pair of opposing ions. 
The key observation is that, in this way, rows corresponding to ions from different pairs remain orthogonal, such that all interactions between these ions are canceled. 
In contrast, opposing ions forming a single pair are assigned identical pulse patterns, such that their mutual interaction never changes sign and is preserved throughout the DD sequence, $\widetilde{J}_{i,i+n}^{\mathrm{eff}}=J_{i,i+n}^{\mathrm{eff}}$.
In the following, we only consider the sign matrix with $n$ rows that describes a single ion chain, keeping in mind that each row defines the pulse sequence for a pair of opposing ions and thus fully specifies the decoupling sequence for the entire double-well system.

Although the concrete assignment of the pulse sequences to the individual ions does not affect the decoupling itself,  a specific ordering simplifies the subsequent extension of the decoupling scheme to account for the spatial dependence of the interactions.
We therefore assign the pulse sequences with the largest number of pulses to the ions in the center of the chains, while the pulse sequences with the smallest number of pulses are assigned to the ions at the edges, such that the two edge ions receive zero and one sign change, respectively. 

An overview of the construction of the decoupling scheme is shown in Fig.~\ref{fig:decoupling_scheme}(a).
As an illustrative example, we consider a chain of $n=6$ ions.
Following the procedure described above, we first construct $\mathcal{H}_8$ recursively from $\mathcal{H}_4$~\eqref{eqn:S4} using Eq.~\eqref{eqn:H_construction}, order the rows as described above, and subsequently remove the two central rows corresponding to $W_6$ and $W_7$ to obtain
\begin{equation}\label{eqn:S6}
    {\setlength{\arraycolsep}{2.5pt}
    \mathcal{S}_6\!=\!\!\begin{pmatrix}
        +1 & +1 & +1 & +1 & +1 & +1 & +1 & +1\\
        +1 & +1 & -1 & -1 & -1 & -1 & +1 & +1\\
        +1 & -1 & -1 & +1 & +1 & -1 & -1 & +1\\
        +1 & -1 & -1 & +1 & -1 & +1 & +1 & -1\\
        +1 & +1 & -1 & -1 & +1 & +1 & -1 & -1\\
        +1 & +1 & +1 & +1 & -1 & -1 & -1 & -1
    \end{pmatrix}\!\!=\!\!\begin{pmatrix}W_0\\W_2\\W_4\\W_5\\W_3\\W_1\end{pmatrix}\!\!.}
\end{equation}
The resulting pulse scheme for two parallel ion strings with $n=6$ ions each, which preserves the desired interactions between opposing ions, is illustrated in Fig.~\ref{fig:decoupling_scheme}(b).

\subsubsection{Geometric cancellation}
Besides DD, the spatial dependence of $J_{ij}^{\mathrm{eff}}$ in the effective Hamiltonian~\eqref{eqn:effective_hamiltonian} provides a complementary way of eliminating selected couplings without additional pulses.
By appropriately choosing the laser wave vector, the interaction between ions $i$ and $j$ can be canceled directly through
\begin{equation}\label{eqn:geometric_cancellation}
\begin{aligned}
J_{ij}^\mathrm{eff}&=\chi_{ij}\cos\bigl(k_\mathrm{L}^z r_{ij,z}^{(0)}\bigr)\stackrel{!}{=}0,\\[5pt]
\Rightarrow\quad
|r_{ij,z}^{(0)}|k_\mathrm{L}^z&=\frac{\pi}{2}(2p+1),\quad p\in\mathds{N}_0.
\end{aligned}
\end{equation}
Here, we have used that the equilibrium ion positions have a negligible $y$-component, such that for a laser wave vector in the $y$-$z$ plane the spatial phase reduces to ${\mathbf{k}_\mathrm{L}\!\cdot\mathbf{r}_{ij}^{(0)}=k_\mathrm{L}^z r_{ij,z}^{(0)}}$ and thus depends only on the inter-ion separation along the direction of the ion chains.
The corresponding coupling can therefore be suppressed by fulfilling a specific relation between this separation and the laser wave vector, as given in Eq.~\eqref{eqn:geometric_cancellation}.
The same choice simultaneously suppresses couplings between ion pairs separated by odd multiples of $r_{ij,z}^{(0)}$.
In particular, the interaction between ions $u$ and $v$ also vanishes if $|r_{uv,z}^{(0)}|=(2q+1)|r_{ij,z}^{(0)}|$ with $q\in\mathds{N}_0$.

For $n>2$, geometric cancellation (GC) alone is generally not sufficient to generate the complete transversal interaction pattern.
However, it can be combined efficiently with the DD scheme introduced above, since interactions that are canceled geometrically no longer need to be removed by the pulse sequence.
To illustrate this, we return to the example of $n=6$ qubits, for which the DD sequence is described by the sign matrix in Eq.~\eqref{eqn:S6}.
When GC is included, instead of starting from $\mathcal{S}_8$ as in the purely DD-based construction, the corresponding decoupling sequence can be constructed starting from $\mathcal{S}_4$ (cf. Eq.~\eqref{eqn:S4}) and duplicating its first and last rows.
The resulting sign matrix then reads
\begin{equation}\label{eqn:S6_GC}
   \mathcal{S}_6^{\mathrm{GC}} =\begin{pmatrix}
        +1&+1&+1&+1\\
        +1&+1&+1&+1\\
        +1&-1&-1&+1\\
        +1&-1&+1&-1\\
        +1&+1&-1&-1\\
        +1&+1&-1&-1
    \end{pmatrix}
    =\begin{pmatrix}W_0\\W_0\\W_2\\W_3\\W_1\\W_1\end{pmatrix},
\end{equation}
which provides a more compact DD synthesis of the desired transversal gate.
Note that, before duplicating, we reorder the rows according to the number of sign changes of the corresponding Walsh functions, as discussed in the previous section.
While the duplicated rows are not orthogonal and therefore do not refocus the couplings between the corresponding edge qubits, these remaining interactions are suppressed by GC.
The orthogonality of all remaining row pairs ensures that all other non-transversal interactions are decoupled.
Although the ions within each string are not equally spaced, with smaller separations near the center than at the edges, the qubits participating in the remaining edge couplings have identical axial separations due to the mirror symmetry of each string.
This allows them to be decoupled simultaneously by choosing $k_\mathrm{L}$ as defined in Eq.~\eqref{eqn:geometric_cancellation} for $i=1$ and $j=2$.
This also explains why it is advantageous to assign the Walsh functions with the fewest sign changes to the first and last row:
Duplicating these rows introduces two additional qubits while increasing the pulse count by only one.
For the $n=6$ example, the combined scheme requires only $4$ time intervals instead of $8$ and reduces the pulse count per well from $15$ to $7$ compared with the purely DD-based construction in Eq.~\eqref{eqn:S6}.
The resulting decoupling pulse sequences without and with GC are illustrated in Figs.~\ref{fig:decoupling_scheme}(b) and (c), respectively.

The construction of $\mathcal{S}_n^\mathrm{GC}$ can be extended straightforwardly to arbitrary~$n\geq 4$ and is illustrated in Fig.~\ref{fig:decoupling_scheme}(a).
We first construct the decoupling sign matrix $\mathcal{S}_{n-2}$ following the procedure introduced in the previous section, with its rows ordered such that the first and last rows contain the fewest sign changes.
To obtain $\mathcal{S}_n^{\mathrm{GC}}$, we duplicate these two rows, thereby adding two qubits to the scheme.
The duplicated rows leave only the two nearest-neighbor edge couplings uncanceled by the DD sequence, while all remaining unwanted interactions are suppressed by the orthogonality of the corresponding rows.
These residual edge couplings are subsequently eliminated geometrically by choosing $k_\mathrm{L}^z|r_{1\,2,z}^{(0)}|=\pi(2p+1)/2$ with $p\in\mathds{N}_0$.
As discussed before, the pulse sequence specified by $\mathcal{S}_n^\mathrm{GC}$ is applied to the full double-well system by performing each $X$-pulse simultaneously on the corresponding pair of opposing ions, $i$ and $i+n$.

The cases with $n<4$ must be treated separately. 
For $n=2$, the single unwanted coupling can be removed either directly by GC or by applying a single decoupling pulse, such that combining both methods provides no additional benefit. 
For $n=3$, GC is used to suppress the nearest-neighbor couplings, while the remaining next-nearest-neighbor interaction is canceled by applying a single $X$-pulse to one of the edge qubits halfway through the evolution time.

\subsection{Gate parameters}\label{sec:gate_parameters}
Having established a framework for engineering the interaction pattern required for the parallel realization of a transversal \CZ gate, we now turn to the choice of parameters governing the resulting gate dynamics.
Beyond the structure of the effective interactions, its realization is determined by a number of parameters that influence both the gate dynamics and its achievable performance.
In the following, we discuss the choice of detuning, gate duration, and Rabi frequency, and introduce the fidelity measure used to characterize the implemented gate.

\begin{figure*}[t]
    \centering
    \includegraphics[scale=1]{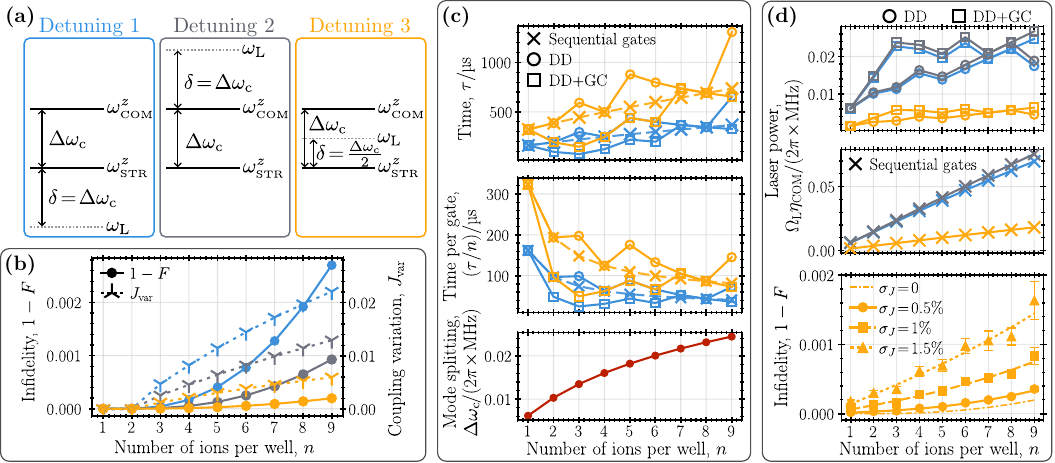}
    \caption{
    Performance, gate duration, and required Rabi frequency of the parallel transversal entangling gate.
    \textbf{(a)}~The three detuning scenarios considered in this work defined by the position of the laser beatnote frequency $\omega_\mathrm{L}$ relative to the axial STR and COM modes.
    \textbf{(b)}~Intrinsic infidelity and coupling variation $J_\mathrm{var}$~\eqref{eqn:coupling_variance} versus the number of ions per well $n$. The infidelity originates from nonuniform STR-mode amplitudes, which lead to unequal interaction strengths and entangling-phase errors. 
    This effect is strongest in scenario~1 due to the larger relative contribution of the STR mode, and smallest in scenario~3.
    \textbf{(c)}~Total gate time (top) and time per entangling operation (middle) for the parallel implementations with and without GC, compared with sequential gates.
    Only scenarios~1 and~3 are shown because scenarios~1 and~2 yield identical gate times. The bottom panel shows the STR--COM mode splitting $\Delta\omega_\mathrm{c}$, which sets the phase-space cycle time.
    \textbf{(d)}~Required Rabi frequency~$\Omega_\mathrm{L}\eta_\mathrm{COM}$ for the parallel schemes (top) and sequential gates (middle). The parallel implementation requires a Rabi frequency smaller by a factor $\sqrt{k}$ for $k$ phase-space cycles. 
    The bottom panel shows the infidelity in detuning scenario~3 for quasi-static entangling pulse-area fluctuations with relative standard deviation $\sigma_J$, which can arise, e.g., from fluctuations in the laser intensity.
    The data points are obtained from Monte Carlo simulations, while the lines show the analytical estimate obtained by combining the intrinsic fidelity (cf. panel~(b)) with the fluctuation-induced fidelity in Eq.~\eqref{eqn:fluctuation_fidelity}.
    For $\sigma_J=0$, only the intrinsic infidelity remains.
    Error bars not visible are smaller than the marker size.
}
\label{fig:results_fidelity}
\end{figure*}

As discussed in the derivation of the Hamiltonian in Sec.~\ref{sec:Hamiltonian}, we assume that the laser frequency is close to the axial COM and STR mode frequencies, so that the detunings $\delta_\mathrm{STR}$ and $\delta_\mathrm{COM}$ are small.
We consider three different choices of the laser frequency $\omega_\mathrm{L}$ relative to the two modes, as illustrated in Fig.~\ref{fig:results_fidelity}(a). 
In the first scenario, the laser frequency is chosen below the STR mode frequency, and we define $\delta\equiv\delta_\mathrm{STR}=\Delta\omega_\mathrm{c}$, which implies $\delta_\mathrm{COM}=2\delta$. 
Conversely, in the second scenario, $\omega_\mathrm{L}$ is placed above the COM mode frequency, and we define $\delta\equiv|\delta_\mathrm{COM}|=\Delta\omega_\mathrm{c}$, such that $|\delta_\mathrm{STR}|=2\delta$. 
Finally, in the third scenario, the laser frequency is chosen symmetrically between the STR and COM modes, such that the two detunings have equal magnitude but opposite sign, $\delta_\mathrm{COM}=-\delta_\mathrm{STR}=\delta$, with $\delta\equiv\Delta\omega_\mathrm{c}/2$. 
These three choices provide the detuning configurations considered throughout the following analysis.

The coupling to the COM and STR modes does not only mediate the effective qubit--qubit interaction, but also drives state-dependent phase-space trajectories of these modes during the gate evolution. 
This generally leads to entanglement between the internal and vibrational degrees of freedom at intermediate times. 
A crucial requirement for the gate implementation is therefore that the phase-space trajectories of all driven modes are closed at the end of the gate, such that the vibrational degrees of freedom disentangle from the qubits.
As derived in Appendix~\ref{app:magnus-expansion}, a mode driven with detuning $\delta_m$ completes one full trajectory in phase space after a time $\tau_m=2\pi/|\delta_m|$. 
To ensure that the trajectories of both the STR and COM modes are closed at the end of the gate, the gate duration must be chosen as an integer multiple of the common closure time,
\begin{equation}\label{eqn:gate_time_ps}
    \tau=\frac{2\pi k}{\delta},\qquad k\in\mathds{N},
\end{equation}
where $\delta>0$ is defined according to the respective detuning scenario. 
For the symmetric choice in scenario~3, this means both modes complete $k$ phase-space cycles during the gate. 
In scenarios~1 and~2, the magnitude of one mode detuning is twice that of the other. 
As a consequence, in scenario~1 the STR and COM modes complete $k$ and $2k$ cycles, respectively, and vice versa in scenario~2.

Since the phase-space trajectories are state dependent, the $X$-pulses applied during the decoupling sequence modify their evolution. 
Consequently, choosing the total gate duration according to Eq.~\eqref{eqn:gate_time_ps} alone no longer guarantees that all trajectories close at the end of the gate when decoupling pulses are applied at intermediate times. 
In standard trapped-ion entangling gates, it is widely studied how to suppress residual qubit--motion entanglement by shaping the state-dependent drive such that the phase-space displacements of the relevant motional modes vanish at the end of an operation. 
This can be achieved, for example, through amplitude, frequency, or phase modulation of the driving field~\cite{choi2014, green2015, leung2018, bluemel2021}. 
Applying such techniques in the present setting would require a separate adaptation to each specific decoupling sequence, as both the pulse pattern and timing depend on the system size and on whether GC is employed.
Moreover, a generic modulation of the state-dependent drive also changes the interaction during the individual decoupling intervals and can therefore interfere with the cancellation condition underlying the sign-matrix construction. 
Hence, we do not employ any additional trajectory shaping technique, but instead apply the decoupling $X$-pulses only when the driven mode trajectories are closed, at the expense of a longer gate time.
We assume the $X$-pulses to be effectively instantaneous on the timescale of the spin--motion dynamics~\cite{harty2016, leu2023}, with their duration $\tau_X$ satisfying $\tau_X|\delta_m|\ll1$ and $\tau_X|g_{i,m}|\ll1$ for all ions $i$ and both modes $m\in\{\mathrm{COM},\mathrm{STR}\}$.
In this way, the qubit and motional degrees of freedom are disentangled whenever the internal qubit states are changed, while phase-space closure at the end of the gate is retained. 
Since the decoupling sequence consists of equal time intervals, we choose the number of phase-space cycles $k$ equal to the number of intervals $n'$ in the respective decoupling sequence (cf. Fig.~\ref{fig:decoupling_scheme}(a)). 
This gives $k=2^{\lceil\log_2(n)\rceil}$ for the purely DD scheme, while including GC yields $k=2^{\lceil\log_2(n-2)\rceil}$ for $n\geq4$, with $k=1$ and $k=2$ for $n=2$ and $n=3$, respectively.
Accordingly, each decoupling interval contains one complete phase-space cycle of both modes in scenario~3, while in scenarios~1 and~2 the mode with the smaller detuning completes one cycle and the other mode two cycles.

As discussed at the beginning of Sec.~\ref{sec:Interaction_Engineering}, an evolution time $\tau=\pi/(8|J|)$ is required to accumulate the entangling phase of a \CZ gate between each pair of opposing ions. 
The gate duration must satisfy both this phase condition and the phase-space closure condition in Eq.~\eqref{eqn:gate_time_ps}, such that $\tau=\pi/(8|J|)=2\pi k/\delta$. 
Assuming uniform mode amplitudes, $\mathcal{M}_{\mathrm{COM}}=(\mathbf{1}_n,\mathbf{1}_n)^{\mathrm{T}}/\sqrt{N}$ and $\mathcal{M}_{\mathrm{STR}}=(\mathbf{1}_n,-\mathbf{1}_n)^{\mathrm{T}}/\sqrt{N}$, where $\mathbf{1}_n=(1,\ldots,1)^{\mathrm{T}}$ denotes the $n$-dimensional all-ones vector, the interaction strength is identical for all pairs of opposing ions and is given by
\begin{equation}\label{eqn:J_approx}
    |J|=\frac{\Omega_\mathrm{L}^2}{4N}\bigg|\frac{\eta_\mathrm{STR}^2}{\delta_\mathrm{STR}}-\frac{\eta_\mathrm{COM}^2}{\delta_\mathrm{COM}}\bigg|.
\end{equation}
The two conditions on the gate duration can then be fulfilled simultaneously by choosing the effective two-photon Rabi frequency according to
\begin{equation}\label{eqn:Rabi_frequency}
    \Omega_{\mathrm{L}}^2=\frac{\delta N}{4k}\bigg|\frac{\delta_\mathrm{STR}\delta_\mathrm{COM}}{\eta_\mathrm{STR}^2\delta_\mathrm{COM}-\eta_\mathrm{COM}^2\delta_\mathrm{STR}}\bigg|.
\end{equation}
Notably, the required Rabi frequency decreases with increasing~$k$. 
Since the total entangling phase is accumulated over a larger number of phase-space cycles, a smaller phase needs to be acquired during each individual cycle, such that a lower Rabi frequency is sufficient to reach the required final entangling phase.

The expression for $|J|$ above, and hence the corresponding gate duration, relies on the assumption of uniform STR and COM mode amplitudes. 
However, the normal modes obtained from diagonalizing the Hessian matrix~\eqref{eqn:Hesse-matrix} show that the STR mode exhibits nonuniform amplitudes across the ion chain, with ions near the center showing larger amplitudes than those near the edges.
For this reason, the interaction strengths between opposing ion pairs are not exactly equal, such that the individual pairs accumulate slightly different entangling phases over the course of the gate. 
The resulting phase errors constitute an intrinsic source of infidelity in the transversal gate implementation and are included in our model.
To quantify the variation of the relevant interaction strengths, we define the relative coupling variation
\begin{equation}\label{eqn:coupling_variance}
    J_{\mathrm{var}}=\frac{\operatorname{std}\big(J_{i,i+n}^\mathrm{eff}\big)}{\big|\langle J_{i,i+n}^\mathrm{eff}\rangle\big|},
\end{equation}
where the average and standard deviation are taken over all $n$ pairs of opposing ions.
The corresponding \ZZ-interaction angle accumulated by the opposing pair~${(i,i+n)}$ during the gate is $\varphi_i=2J_{i,i+n}^\mathrm{eff}\tau$, while the ideal gate requires $|\varphi|=\pi/4$.
Writing $J_{i,i+n}^\mathrm{eff}=\langle J_{i,i+n}^\mathrm{eff}\rangle+\Delta J_i$ and assuming $|\langle J_{i,i+n}^\mathrm{eff}\rangle|\approx|J|$, we obtain
\begin{equation}\label{eqn:phase_variation}
    |\varphi_i|\approx\frac{\pi}{4}\left(1+\frac{\Delta J_i}{\langle J_{i,i+n}^\mathrm{eff}\rangle}\right),
\end{equation}
showing that the relative coupling variation directly determines the spread of the accumulated \ZZ-interaction angles around their ideal value.
A more detailed analytical characterization of this effect would require resolving the full mode structure and interaction pattern of the double-well system.
Instead, we quantify the impact of this error source on the gate performance numerically below.

To analyze the performance of the transversal \CZ gate, we consider its ability to generate a maximally entangled state from the initial product state ${\ket{\Psi_0}=\ket{+}^{\otimes 2n}}$. 
Under an ideal transversal evolution $U_\mathrm{T}=\prod_{i=1}^n \exp(-\ci\,\mathrm{sign}(J)\frac{\pi}{4}Z_iZ_{i+n})$ the resulting target state is
\begin{equation}\label{eqn:target_state}
    \ket{\Psi_\mathrm{T}} = \prod_{i=1}^n\frac{1}{\sqrt{2}}\Big(\!\ket{+}_i\!\ket{+}_{i+n}\!-\ci\,\mathrm{sign}(J)\!\ket{-}_i\!\ket{-}_{i+n}\!\Big).
\end{equation}
We use the fidelity with respect to this maximally entangled target state as a measure of the quality of the implemented transversal gate,
\begin{equation}\label{eqn:fidelity}
    F=\big|\!\bra{\Psi_0}U_\mathrm{T}^\dagger U_\mathrm{eff}\ket{\Psi_0}\!\big|^2,
\end{equation}
where $U_\mathrm{eff}$ denotes the evolution generated by the engineered effective interaction.

\section{Numerical gate performance}\label{sec:results}
For the following simulations, we use the trap parameters $\omega_x=\freq{2.0}$, $\omega_y=\freq{2.1}$, $\omega_z=\freq{0.48}$, and $d=\SI{31}{\micro\meter}$, motivated by experimental double-well $^{40}\text{Ca}^+$ platforms~\cite{valentini2025}.
For each system size, we numerically compute the interaction matrix $J_{ij}^\mathrm{eff}$ and simulate the time evolution of the initial state $\ket{\Psi_0}$ under the resulting Hamiltonian~\eqref{eqn:effective_hamiltonian}.
The gate time and Rabi frequency are chosen according to Eqs.~\eqref{eqn:gate_time_ps} and~\eqref{eqn:Rabi_frequency}, with the number of phase-space cycles $k$ fixed by the number of intervals in the respective decoupling sequence.
As discussed in Sec.~\ref{sec:gate_parameters}, throughout the simulations we assume instantaneous decoupling pulses that are applied only when all driven phase-space trajectories are closed, i.e., after each complete cycle of duration $2\pi/\delta$, where $\delta$ is defined according to the respective detuning scenario.

\subsection{Fidelity}
We first revisit the example of $n=6$ ions per well from Sec.~\ref{sec:Interaction_Engineering}, with the corresponding decoupling schemes without and with GC shown in Figs.~\ref{fig:decoupling_scheme}(b) and (c), respectively.
For this example, the purely DD sequence requires $k=8$ phase-space cycles, whereas including GC reduces this to $k=4$, thereby halving the gate duration.
The resulting fidelities, as defined in Eq.~\eqref{eqn:fidelity}, are shown in the lower panels of Figs.~\ref{fig:decoupling_scheme}(b) and (c).
Since these fidelities are obtained within the effective spin model, they do not capture transient spin--motion entanglement at intermediate times, while agreeing with the full two-mode dynamics at the phase-space closure times.
All three detuning scenarios reach a final infidelity below $10^{-3}$ both with and without GC, demonstrating the functionality of the introduced decoupling scheme.

Notably, for each detuning scenario, the final fidelity is the same with and without GC despite the different gate durations.
This follows from the fact that the dominant error source in our model is the variation of the interaction strengths between opposing ion pairs, which causes the pairs to accumulate slightly different entangling phases.
Adjusting $\Omega_\mathrm{L}$ according to Eq.~\eqref{eqn:Rabi_frequency} compensates for the different gate durations, such that these phase errors are independent of the number of phase-space cycles $k$ and hence of whether GC is applied.
The final fidelity, however, depends on the chosen detuning scenario.
To investigate this dependence across different system sizes, Fig.~\ref{fig:results_fidelity}(b) shows the infidelity and the corresponding relative coupling variation $J_\mathrm{var}$ (cf. Eq.~\eqref{eqn:coupling_variance}) for $n\in[1,9]$.
For all system sizes, scenario~3 gives the lowest infidelity and coupling variation, while scenario 1 gives the highest, with the differences becoming more pronounced as the system size increases.

The differences between the three detuning scenarios can be understood from the relative contributions of the COM and STR modes to the effective interaction.
While the uniform COM mode does not introduce variations between opposing ion pairs, the nonuniform STR mode is responsible for the coupling variation and, according to Eq.~\eqref{eqn:phase_variation}, for the resulting phase errors between the individual ion pairs.
Because scenario 1 couples more strongly to the STR mode than scenario 2, it exhibits a larger $J_\mathrm{var}$ and correspondingly larger phase errors.
A particularly small variation is obtained for scenario 3, where the laser frequency lies symmetrically between the two modes.
Here, the COM and STR detunings have opposite signs and, together with the opposite parity of the mode eigenvectors between the wells, cause the two contributions to add constructively rather than partially cancel as in scenarios 1 and 2~\cite{valentini2025}.
Since only the STR contribution varies between ion pairs, this constructive addition increases the mean interaction strength relative to its variation and thereby reduces both $J_\mathrm{var}$ and the corresponding phase errors.
Consequently, detuning scenario 3 provides the best gate performance in terms of the intrinsic infidelity considered here.

\subsection{Gate time}
The upper panel in Fig.~\ref{fig:results_fidelity}(c) compares the gate durations of the parallel implementations with and without GC with a sequential implementation for different system sizes $n\in[1,9]$.
Since detuning scenarios 1 and 2 have the same detuning $\delta$ and therefore yield identical gate times for all $n$, we only show scenario~1 together with scenario~3.
For the sequential implementation, we assume that the $n$ opposing ion pairs are addressed individually and entangled one after another, with each two-qubit gate requiring a single phase-space cycle such that the total sequential gate time is $\tau_\mathrm{S}=2\pi n/\delta$.
As expected, the results confirm that the larger detuning in scenario~1, which is twice that of scenario~3, reduces the gate time by a factor of two.
Note that this benchmark compares coherent interaction times and omits transport, finite-duration control pulses, and calibration overheads.

The system-size dependence of the gate duration ${\tau=2\pi k/\delta}$ is determined by both the number of phase-space cycles $k$ and the detuning $\delta$. 
The latter is set by the splitting $\Delta\omega_\mathrm{c}$ between the STR and COM modes, shown in the lower panel of Fig.~\ref{fig:results_fidelity}(c). 
As $\Delta\omega_\mathrm{c}$ generally increases with the system size, the corresponding phase-space cycle time decreases with increasing $n$.
This effect becomes more apparent in the central panel of Fig.~\ref{fig:results_fidelity}(c), which shows the gate time per entangling operation. 
Overall, this quantity decreases with increasing system size, demonstrating that the gate execution per entangled ion pair becomes faster for larger systems due to increasing detuning $\delta\propto\Delta\omega_\mathrm{c}$.

In addition to this, the number of required phase-space cycles introduces a stepwise dependence of the gate duration on the system size. 
For the parallel implementation using only DD, the number of cycles is $k=2^{\lceil\log_2(n)\rceil}$ and therefore increases whenever the next power of two is exceeded, resulting in corresponding jumps in the gate duration. 
Including GC reduces the required number of cycles to $k=2^{\lceil\log_2(n-2)\rceil}$ for $n\geq4$, and to $k=1$ and $k=2$ for $n=2$ and $n=3$, respectively, thereby shifting the jumps by two ions and generally shortening the gate duration.
As a result, including GC makes the parallel implementation faster than or equally fast as the sequential approach for most system sizes shown in Fig.~\ref{fig:results_fidelity}(c), with $n=7$ being the only exception.
More generally, a parallel implementation provides a speedup whenever $k<n$. 
This condition is never satisfied for the purely DD-based scheme. For the combined scheme and $n\geq4$, it holds whenever $n-1$ or $n-2$ is a power of two, corresponding to $n=4,5,6,9,10,17\ldots$. 
The special cases $n=2$ and $n=3$ discussed in Sec.~\ref{sec:Interaction_Engineering} also provide a speedup, requiring only $k=1$ and $k=2$ phase-space cycles, respectively.

\subsection{Laser power}
As discussed in the previous section, we choose the Rabi frequency according to Eq.~\eqref{eqn:Rabi_frequency} such that the required entangling phase is accumulated within the available gate time. 
The required Rabi frequency depends both on the chosen detuning scenario and on the number of phase-space cycles $k$, which in turn is determined by the employed decoupling scheme. 
To quantify the corresponding Rabi-frequency requirements, the upper panel of Fig.~\ref{fig:results_fidelity}(d) shows $\Omega_\mathrm{L}\eta_\mathrm{COM}$ as a function of the system size for the parallel implementations with and without GC, while the central panel shows the corresponding values for sequential two-qubit gates.

Comparing the upper and central panels of Fig.~\ref{fig:results_fidelity}(d) shows that the sequential implementation requires a substantially higher Rabi frequency than the parallel schemes. 
For the sequential implementation, we assume that each two-qubit gate is completed within a single phase-space cycle, such that the full entangling phase must be accumulated during that time. 
In the parallel implementation, by contrast, the required phase is accumulated over $k$ phase-space cycles, allowing for a considerably smaller Rabi frequency.
Accordingly, the sequential implementation requires a Rabi frequency that is larger by a factor of $\sqrt{k}$ compared with the parallel implementation.
The required Rabi frequency of the sequential gates could be reduced by increasing the number of phase-space cycles per two-qubit gate, but only at the cost of proportionally increasing the total execution time. 
Thus, the parallel implementation allows multiple ion pairs to be entangled simultaneously while requiring a substantially lower two-photon Rabi frequency, making it particularly favorable from an experimental perspective, especially when scaling to larger system sizes.

Within the parallel implementations, the required Rabi frequency further depends on the employed decoupling scheme.
For a fixed detuning scenario, including GC leads to a larger or equal required Rabi frequency due to the smaller or equal number of phase-space cycles during the gate.
Since fewer cycles are available, a larger phase has to be acquired during each cycle and hence the Rabi frequency needs to be larger.
Thus, the reduction in gate duration achieved by GC comes at the cost of a higher Rabi frequency.
A pronounced difference is also observed between the detuning scenarios, with scenario~3 requiring a substantially smaller effective two-photon Rabi frequency than scenarios~1 and~2. 
This can be understood from the contributions of the STR and COM modes to the effective interaction in Eq.~\eqref{eqn:J_approx}. 
In scenario~3, the opposite signs of the two detunings cause the two mode contributions to add constructively, whereas they partially cancel in scenarios~1 and~2. 
For this reason, a given Rabi frequency produces a stronger effective interaction in scenario~3, such that a smaller $\Omega_\mathrm{L}$ is sufficient to accumulate the required entangling phase within the prescribed gate duration.

\section{Error analysis}\label{sec:error_analysis}
In this section, we assess the robustness of the transversal gate against experimentally relevant imperfections that are not included in the idealized analysis above.
We consider entangling pulse-area fluctuations, residual coupling to spectator modes, qubit dephasing, and wave-vector misalignment, and analyze how these error sources affect the gate performance.
Unless stated otherwise, we use the same trap parameters and gate settings as in the previous section.

\subsection{Entangling pulse-area fluctuations}
Experimental imperfections can lead to fluctuations in the entangling pulse area and thereby result in deviations of the accumulated interaction phase from its target value.
Such fluctuations can arise from different microscopic error sources, including fluctuations of the laser intensity or the detuning.
Rather than modeling these individual sources explicitly, we account for their effect directly at the level of the effective interaction strengths by considering $J_{ij}^{\mathrm{eff}}\rightarrow J_{ij}^{\mathrm{eff}}(1+\epsilon_J)$, where $\epsilon_J$ denotes the relative interaction-strength deviation.
For a fixed gate duration, $\epsilon_J$ directly corresponds to the relative deviation of the entangling pulse area.
To quantify this effect, we perform a Monte Carlo simulation assuming quasi-static pulse-area fluctuations, modeled by a common relative deviation $\epsilon_J$ of all interaction strengths that varies between individual runs but remains constant throughout each simulation.

For each realization, we sample $\epsilon_J$ from a normal distribution $\mathcal{N}(0,\sigma_J^2)$ with standard deviation $\sigma_J$ and simulate the gate using the resulting interaction strengths.
Considering detuning scenario~3 and system sizes $n\in[1,9]$, we perform $N_\mathrm{R}=100$ gate simulations for each data point.
The resulting infidelities for $\sigma_J\in\{0.5,1.0,1.5\}\%$ are shown in the lower panel of Fig.~\ref{fig:results_fidelity}(d).
We also show the intrinsic infidelity at $\sigma_J=0$, which arises from variations in the interaction strengths due to the nonuniform STR-mode amplitudes, as discussed above.
Each data point corresponds to the mean fidelity
\begin{equation}
    \bar{F}=\frac{1}{N_\mathrm{R}}\sum_{r=1}^{N_\mathrm{R}}F_r,
\end{equation}
while the error bars indicate the Monte Carlo standard error (MCSE) $s_\mathrm{F}/\sqrt{N_\mathrm{R}}$, with
\begin{equation}
    s_\mathrm{F}=\sqrt{\frac{1}{N_\mathrm{R}-1}\sum_{r=1}^{N_\mathrm{R}}(F_r-\bar{F})^2}
\end{equation}
denoting the sample standard deviation.
As expected, the infidelity increases with increasing pulse-area fluctuations.
As derived in App.~\ref{app:fluctuation_error}, for small fluctuations, $|\epsilon_J|\ll1$, and in the absence of intrinsic errors, the expected fidelity is given by
\begin{equation}\label{eqn:fluctuation_fidelity}
    \bar{F}_\mathrm{fluct}\approx 1-\frac{n\pi^2}{16}\sigma_J^2.
\end{equation}
Assuming that the intrinsic and fluctuation-induced errors are independent, combining the two contributions yields a total infidelity of approximately $1-F_\mathrm{intr}\bar{F}_\mathrm{fluct}$.
The corresponding analytical predictions are shown by the lines in the lower panel of Fig.~\ref{fig:results_fidelity}(d) and agree well with the Monte Carlo results.
Hence, within the quasi-static fluctuation model considered here, the infidelity caused by entangling pulse-area fluctuations scales linearly with the number of entangled ion pairs $n$ and quadratically with the relative fluctuation~$\sigma_J$.
Notably, the sensitivity of the gate to relative pulse-area fluctuations does not depend on the chosen detuning scenario or the decoupling scheme.
Thus, the same fluctuation-induced error is expected for parallel and sequential implementations of the entangling operations for a given relative pulse-area stability.

\subsection{Coupling to spectator modes}\label{sec:spectator_mode_coupling}
In the derivation of the effective interaction Hamiltonian in Sec.~\ref{sec:Hamiltonian}, we have neglected the coupling to the spectator modes, assuming their detunings from the laser beatnote to be large compared with those of the targeted COM and STR modes.
In practice, however, the finite off-resonant coupling to these modes gives rise to small state-dependent displacements in phase space.
While these displacements remain bounded during an uninterrupted gate evolution, the intermediate $X$-pulses of the decoupling sequence can modify their accumulation.
As discussed in Sec.~\ref{sec:gate_parameters}, we avoid disturbing the phase-space trajectories of the COM and STR modes by applying the $X$-pulses only when both trajectories are closed.
The phase-space trajectories of the spectator modes, however, are not necessarily closed whenever the decoupling pulses are applied, and the applied sequence might cause these trajectories to drift far from the origin, resulting in residual entanglement between motional and qubit degrees of freedom. 
In this section, we quantify the spectator-mode contribution to the infidelity of the parallel entangling operation.

Since the chosen laser geometry does not couple to the $x$-modes and we assume the radial $y$-modes to be sufficiently far detuned from the laser beatnote, we restrict the following analysis to the axial spectator modes.
We therefore explicitly include all $N=2n$ axial $z$-modes, comprising the targeted COM and STR modes as well as the remaining $N-2$ axial spectator modes.

We evaluate the resulting fidelity using the exact analytical expression derived in App.~\ref{app:spectator_modes}, which accounts for the free evolution during each decoupling interval and the intermediate $X$-pulses.
For weak spectator-mode coupling, corresponding to small phase-space displacements, the fidelity can be approximated as
\begin{equation}\label{eq_fidelity_spectator_modes_approximation}
    F \approx \prod_{i=1}^{n} \cos^2(\Delta \varphi_i) - \sum_{m=1}^{2n} (2\overline{n}_m+1) \sum_{j=1}^{2n} |\tilde{\alpha}_{j,m}|^2,
\end{equation}
where $\overline{n}_m$ denotes the average phonon occupation of mode $m$, and $\Delta\varphi_i$ quantifies the deviation from the ideal entangling phase of the opposing ion pair $i,i+n$ 
arising from the finite off-resonant coupling to the spectator modes and the nonuniform interaction strengths between opposing ion pairs.
Therefore, the intrinsic phase errors discussed in Sec.~\ref{sec:gate_parameters} are retained throughout this analysis and contribute to the reported infidelities.
The spectator-mode contribution to $\Delta\varphi_i$ could in principle be accounted for when experimentally calibrating the $ZZ$ gate duration, but is treated here as an additional source of error.
The residual displacement of mode $m$ induced by ion $j$ after the complete decoupling sequence is described by
\begin{equation}\label{eq_accumulated_displacements_spectator_modes}
    \tilde{\alpha}_{j,m} 
    = \sum_{l=1}^{k}\! \alpha_{j,m}^{(l)} \mathcal{S}_{jl} 
    =  \frac{g_{j,m}}{\delta_m} \sum_{l=1}^{k}  (e^{\ci \delta_m t_{l-1}} - e^{\ci \delta_m t_l}) \mathcal{S}_{jl},
\end{equation} 
where $\alpha_{j,m}^{(l)}$ denotes the displacement contribution accumulated during the $l$th decoupling interval, $g_{j,m}$ is the ion--mode coupling introduced together with the dipole Hamiltonian in Eq.~\eqref{eqn:dipole_Hamiltonian_LDA}, and $\delta_m$ is the corresponding mode detuning.
The sign matrix $\mathcal{S}$ describes the decoupling sequence introduced in Sec.~\ref{sec:Interaction_Engineering}, where we consider $\mathcal{S}_{i+n,l}=\mathcal{S}_{il}$ for $i=1,\dots,n$. Here, $k$ denotes the number of decoupling intervals and $t_l$ the boundary of the $l$th interval.
Note that the residual displacements $\tilde{\alpha}_{j,m}$ vanish for the targeted COM and STR modes, corresponding to $m=1$ and $m=2$, respectively.
A detailed derivation of Eqs.~\eqref{eq_fidelity_spectator_modes_approximation} and~\eqref{eq_accumulated_displacements_spectator_modes} is provided in Appendix~\ref{app:spectator_modes}.

\begin{figure}[t]
    \centering
    \includegraphics[scale=1]{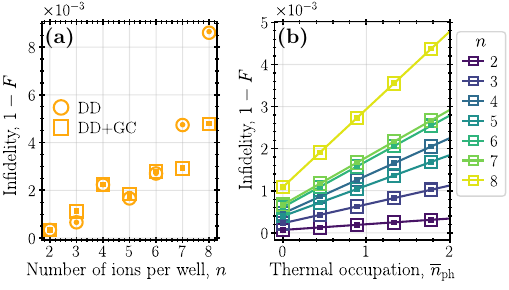}
    \caption{
    Infidelity in the presence of residual coupling to axial spectator modes for detuning scenario~3.
    \textbf{(a)}~Infidelity as a function of the number of ions per well for decoupling sequences with and without GC at a fixed average occupation $\overline{n}_\mathrm{ph}=2$.
    Open markers show the full analytical result, while filled markers correspond to the small-displacement approximation in Eq.~\eqref{eq_fidelity_spectator_modes_approximation}.
    \textbf{(b)}~Infidelity as a function of the average occupation $\overline{n}_\mathrm{ph}$ for different system sizes $n$, considering the decoupling sequence with GC.
}
    \label{fig:fidelity_spectator}
\end{figure}

We analyze the resulting infidelity for decoupling sequences with and without GC for system sizes $n\in[2,8]$.
For simplicity, we restrict the analysis to detuning scenario~3 and assume the same average phonon occupation for all modes, $\overline{n}_m=\overline{n}_\mathrm{ph}$, as could approximately result from sequential resolved-sideband cooling of the individual axial modes.
Figure~\ref{fig:fidelity_spectator}(a) shows the resulting infidelity for $\overline{n}_\mathrm{ph}=2$, obtained from the exact analytical expression (open markers) and from the small-displacement approximation in Eq.~\eqref{eq_fidelity_spectator_modes_approximation} (filled markers).
The approximation closely reproduces the exact results over the entire range of system sizes considered.
Overall, the infidelity tends to increase with system size, consistent with the increasing number of contributing motional modes and the larger phase errors that can accumulate over longer gate durations.
However, the dependence is not strictly monotonic, and no systematic difference is observed between the decoupling sequences with and without GC.
This behavior can be understood in terms of the coherent accumulation of the individual displacement contributions in Eq.~\eqref{eq_accumulated_displacements_spectator_modes}, which depends on both the spectator-mode frequencies and the sign pattern imposed by the decoupling sequence.
Consequently, the residual displacements can add constructively or destructively for individual modes, leading to the variations observed with system size and between the two decoupling schemes.
A more detailed analysis of how the spectator-mode contribution depends on the mode structure and decoupling sequence is left for future work.

To further investigate the effect of the motional state, Fig.~\ref{fig:fidelity_spectator}(b) shows the infidelity in the presence of spectator-mode coupling as a function of the average phonon occupation $\overline{n}_\mathrm{ph}$ for different system sizes, considering the decoupling sequence with GC.
As expected from Eq.~\eqref{eq_fidelity_spectator_modes_approximation}, the infidelity increases linearly with $\overline{n}_\mathrm{ph}$ for each system size, while the slope shows the same non-monotonic system-size dependence observed in Fig.~\ref{fig:fidelity_spectator}(a), with an overall increase toward larger $n$.
Importantly, a finite infidelity remains for ground-state-cooled modes, $\overline{n}_\mathrm{ph}=0$, which is approximately one order of magnitude larger than the intrinsic infidelity in the absence of spectator-mode coupling (cf. Fig.~\ref{fig:results_fidelity}(b)).
This additional error arises from residual spin-dependent displacements, whose contribution remains finite even for ground-state-cooled modes, as well as from spectator-mode-induced phase errors.
Thus, while ground-state cooling cannot completely eliminate the spectator-mode-induced error, the increasing sensitivity to phonon occupation for larger systems highlights the importance of efficiently cooling the spectator modes.

More sophisticated pulse-shaping techniques can be employed to suppress such residual motional displacements \cite{leung2018, bluemel2021}.
Exploring such optimized pulse shapes in combination with the decoupling sequences considered here is beyond the scope of the present analysis and is left for future work.

\subsection{Qubit dephasing}
Limited qubit coherence has been identified as an important error source in experimental inter-well entangling operations~\cite{valentini2025}. 
To quantify its impact on the transversal gate, we model decoherence as local pure dephasing characterized by a coherence time $T_2$. 
The coherent state-vector simulation is therefore extended to a Lindblad master-equation description, where the density matrix $\rho$ evolves according to
\begin{equation}
    \dot{\rho} = -\ci[H_\mathrm{eff},\rho] + \sum_i \Bigl( L_i \rho L_i^\dagger - \frac{1}{2}\{L_i^\dagger L_i,\rho\} \Bigr)
\end{equation}
with local collapse operators $L_i=Z_i/\sqrt{2T_2}$ acting on each ion $i=1,\ldots,N$.
The coherent dynamics are governed by the effective Hamiltonian in Eq.~\eqref{eqn:effective_hamiltonian}, while the collapse operators lead to an exponential decay of the single-qubit coherences as $\exp(-t/T_2)$.
The decoupling sequences introduced in Sec.~\ref{sec:Interaction_Engineering} can be directly incorporated into the density-matrix evolution by applying each $X$-pulse as $\rho\rightarrow X_i\rho X_i$, where we assume the pulses to be instantaneous.
Starting from $\rho_0=\ket{\Psi_0}\!\!\bra{\Psi_0}$, the final gate fidelity is given by $F=\Tr[\ket{\Psi_\mathrm{T}}\!\!\bra{\Psi_\mathrm{T}}\rho(\tau)]$, where the target state is defined in Eq.~\eqref{eqn:target_state}.
Assuming independent dephasing of all qubits and neglecting the intrinsic coherent gate errors, the resulting fidelity is
\begin{equation}\label{eqn:dephasing_fidelity}
    F_\mathrm{deph}=\left(\frac{1+e^{-\tau/T_2}}{2}\right)^{2n}.
\end{equation}

\begin{figure}[t]
    \centering
    \includegraphics[scale=1]{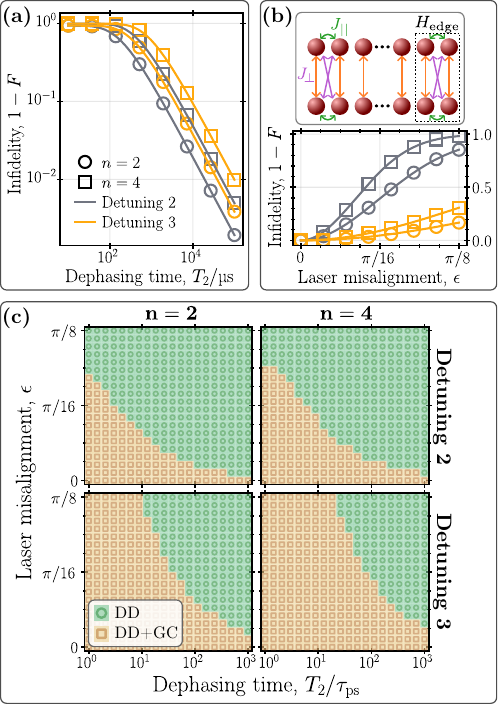}
    \caption{
        Robustness of the transversal entangling gate against qubit dephasing and laser wave-vector misalignment for $n=2,4$ ions per well and detuning scenarios~2 and~3.
        \textbf{(a)}~Gate infidelity versus dephasing time $T_2$ for the implementation including GC. 
        Longer gate durations lead to larger dephasing errors, such that $n=2$ outperforms $n=4$ and scenario~2 outperforms scenario~3.
        \textbf{(b)}~Residual couplings generated by imperfect GC due to wave-vector misalignment $\epsilon$ (top), and the resulting gate infidelity (bottom). 
        The misalignment-induced error is smaller in scenario~3 because the residual couplings are suppressed relative to the desired inter-well interaction (see main text for details).
        \textbf{(c)}~Comparison of the decoupling schemes with and without GC as a function of the wave-vector misalignment~$\epsilon$ and the normalized dephasing time $T_2/\tau_\mathrm{ps}$, with $\tau_\mathrm{ps}=2\pi/\delta$ denoting the duration of one phase-space cycle.
        The preferred scheme is set by the competition between the shorter gate duration obtained with GC, which reduces dephasing errors, and the residual couplings generated by imperfect GC under misalignment.
}
            \label{fig:results_decoherence}
\end{figure}

Figure~\ref{fig:results_decoherence}(a) shows the infidelity for $n=2$ and $n=4$ in detuning scenarios~2 and~3, using the decoupling scheme combined with GC. 
The markers show the numerical master-equation results, while the solid lines correspond to the analytical prediction in Eq.~\eqref{eqn:dephasing_fidelity}. 
Within the model considered here, the dephasing-induced error is determined by the gate duration and the number of dephasing qubits.
Accordingly, scenario~2 performs better than scenario~3 because its gate duration is shorter by a factor of two, while the $n=4$ case exhibits larger errors than $n=2$ due to both the longer gate time and the larger number of ions.
By comparison, the intrinsic infidelities caused by the nonuniform STR-mode amplitudes range from approximately $10^{-8}$ to $10^{-5}$ for the four cases considered here (see Fig.~\ref{fig:results_fidelity}). 
They are therefore several orders of magnitude smaller than the dephasing-induced errors, identifying qubit dephasing as a potentially dominant limitation of the gate performance within the present model.
Note, however, that this model does not account for the potential refocusing of dephasing noise by the applied DD sequences and may therefore overestimate the resulting gate infidelity, as discussed in more detail below.

In an experimental implementation, one contribution to qubit dephasing can arise from fluctuations of the magnetic field, which shift the qubit transition frequencies and thereby lead to the accumulation of relative phases. 
Slowly varying contributions to these fluctuations can be partially refocused by suitable DD pulse sequences~\cite{uhrig2007, biercuk2009, wang2021, weber2024}. 
Although the $X$-pulse sequences employed in our interaction-engineering scheme are designed to cancel unwanted qubit--qubit interactions, they can simultaneously provide such a refocusing effect and thereby partly suppress the magnetic-field-induced dephasing.
In contrast to the equal and fixed $T_2$ assumed for all ions in the master-equation simulations above, this sequence-dependent suppression may result in different effective coherence times for the individual ions.
As shown in App.~\ref{app:magnetic_compensation}, the suppression of dephasing caused by slowly varying magnetic-field fluctuations depends directly on the sign matrix $\mathcal{S}$ of the applied pulse sequence.
For a frequency shift that varies linearly in time during the gate, the constant and linear contributions are completely refocused if
\begin{equation}\label{eqn:magnetic_compensation}
    \sum_{l=1}^{k}\mathcal{S}_{il}=0,
    \qquad
    \sum_{l=1}^{k}(2l-1)\mathcal{S}_{il}=0,
\end{equation}
respectively.
Suppression of the constant contribution is achieved by all Walsh functions except $W_0$, since every non-constant Walsh function contains an equal number of positive and negative entries. 
Suppression of the linear contribution, in contrast, depends on the specific pulse sequence.
For example, for the $n=6$ sequence introduced in Eq.~\eqref{eqn:S6}, both the constant and linear contributions are refocused for ions $2$, $3$, and $4$.
Consequently, the applied decoupling sequences may effectively increase the coherence times of individual ions and thereby reduce the dephasing-induced infidelities compared with the results obtained for the fixed $T_2$ values considered above.

\subsection{Misalignment of the wave vector}\label{sec:misaligned_wave_vector}
In order to reduce the number of pulses in the decoupling scheme and thereby the gate time, we exploit the intrinsic spatial dependence of the interaction strength to geometrically decouple the ions at the edges, as introduced in Sec.~\ref{sec:Interaction_Engineering}.
This requires choosing the laser wave vector according to Eq.~\eqref{eqn:geometric_cancellation} with $i=1$ and $j=2$.
A small misalignment, $|r_{1\,2,z}^{(0)}|k_\mathrm{L}^z =\pi(2p+1)/2+\epsilon$, prevents perfect cancellation and gives rise to residual edge couplings.
As a consequence, the four ions at the edges remain weakly coupled to their neighbors, as illustrated in Fig.~\ref{fig:results_decoherence}(b).
To quantify the resulting error, we first consider a single edge, described by 
\begin{equation}\label{eqn:Hamiltonian_edge}
    \begin{aligned} 
        H_\mathrm{edge} ={}& 2J_{1,1+n}^\mathrm{eff}Z_1Z_{1+n}
        +2J_{2,2+n}^\mathrm{eff}Z_2Z_{2+n}\\
        &+2J_\perp\bigl(Z_1Z_{2+n}+Z_2Z_{1+n}\bigr)\\
        &+2J_{\parallel}\bigl(Z_1Z_2+Z_{1+n}Z_{2+n}\bigr),
    \end{aligned}
\end{equation}
where the first line contains the desired inter-chain interactions, while the remaining terms describe the unwanted residual inter- and intra-chain couplings,
${|J_\perp|=|\chi_{1,2+n}\sin(\epsilon)|}$, and ${|J_{\parallel}|=|\chi_{1,2}\sin(\epsilon)|}$, respectively.
Neglecting intrinsic phase errors in the desired transversal couplings, evaluating the fidelity according to Eq.~\eqref{eqn:fidelity}, with $U_\mathrm{eff}$ given by the time evolution under the edge Hamiltonian~\eqref{eqn:Hamiltonian_edge}, gives
\begin{equation}\label{eqn:fidelity_edge}
    F_\mathrm{edge}=\left[\frac{1+\cos(4\tau J_{\parallel})\cos(4\tau J_\perp)}{2}\right]^2.
\end{equation}
For small misalignments, $|4J_{\parallel/\perp}\tau|\ll1$, the infidelity scales quadratically with the residual coupling strengths and, since $J_{\parallel/\perp}\propto\sin(\epsilon)\simeq\epsilon$, as $\epsilon^2$.

The resulting infidelities are shown in Fig.~\ref{fig:results_decoherence}(b) for $n=2$ and $n=4$ in detuning scenarios~2 and~3, where the data points are obtained from simulations of the full gate dynamics and the solid lines show the analytical predictions based on Eq.~\eqref{eqn:fidelity_edge}.
The larger sensitivity to laser misalignment observed for $n=4$ results from the two spatially separated edge regions at which residual intra- and inter-chain \ZZ couplings remain, whereas for $n=2$ only a single such contribution is present.
Accordingly, we get $F=F_\mathrm{edge}$ for $n=2$ and $F=F_\mathrm{edge}^2$ for $n=4$.
Note that, for $n\geq4$, the number of residual edge couplings remains unchanged, so the misalignment-induced infidelity is independent of the chain length for fixed accumulated phases $\tau J_{\parallel/\perp}$.
Unlike the dephasing-induced infidelity discussed above, the misalignment-induced error is smaller in scenario~3 despite its longer gate duration.
According to Eq.~\eqref{eqn:fidelity_edge}, the error is determined by $\tau J_{\parallel/\perp} = J_{\parallel/\perp}\pi/(8|J|)$ and therefore by the residual edge couplings relative to the desired inter-well interaction~\eqref{eqn:J_approx}.
In scenario~3, the opposite signs of the COM and STR detunings lead to destructive interference of the mode contributions to the residual intra-well coupling $J_\parallel$, while they add constructively for the desired inter-well coupling $J$.
This suppresses $|J_\parallel|/|J|$ and thereby reduces the misalignment-induced infidelity.

The results above reveal a trade-off in the choice of the decoupling scheme when both dephasing and wave-vector misalignment are present.
Including GC reduces the number of required phase-space cycles and hence the gate duration, thereby suppressing dephasing-induced errors.
At the same time, however, imperfect GC introduces the residual \ZZ couplings discussed above, which are absent for the purely DD-based implementation under the assumption of ideal and instantaneous decoupling pulses.
Figure~\ref{fig:results_decoherence}(c) illustrates this trade-off by indicating which of the two schemes yields the higher fidelity as a function of the dephasing time $T_2$ and the laser misalignment $\epsilon$.
For detuning scenario~2, the comparatively strong sensitivity to wave-vector misalignment results in a substantial parameter regime where the implementation without GC achieves a higher fidelity within the error model considered here, particularly at long coherence times where the advantage of the shorter gate duration becomes less relevant.
In contrast, scenario~3 is considerably less sensitive to misalignment, such that the shorter gate duration achieved by including GC reduces dephasing errors over a much larger parameter range.
Thus, whether GC is beneficial ultimately depends on the balance between the coherence time and the magnitude of the wave-vector misalignment, with detuning scenario~3 generally providing the more favorable regime for its application.
We note, however, that this comparison assumes ideal and instantaneous decoupling pulses.
In a realistic implementation, pulse imperfections and finite pulse durations may introduce additional errors, including imperfect refocusing of undesired couplings, whose impact generally increases with the number of applied pulses.
The reduced pulse count achieved by including GC may therefore provide an additional advantage that is not captured by the present analysis.

Beyond their impact on the physical gate fidelity, residual interactions resulting from an incomplete elimination of undesired couplings are particularly relevant in the context of FT quantum computation.
As discussed above, such residual couplings may arise, for instance, from wave-vector misalignment in the GC scheme and, by coupling multiple qubits within the same code block, may compromise the FT properties of the otherwise transversal operation.
In the following section, we therefore investigate to what extent the residual interactions induced by wave-vector misalignment affect the FT properties of the transversal gate.

\section{Impact of crosstalk on fault-tolerant quantum computation}\label{sec:ft_crosstalk}
A key feature of transversal entangling gates between two logical qubits is that they act independently on corresponding physical qubits, thereby preventing a single fault from spreading to multiple qubits within the same code block.
The residual \ZZ interactions induced by wave-vector misalignment can violate this property by introducing additional couplings beyond the intended transversal interaction pattern.
In particular, couplings between qubits within the same code block can induce correlated phase-flip errors and thereby compromise the FT character of the transversal gate.
In the following, we refer to these additional interactions as crosstalk errors.
To assess their impact in an FT setting, we consider the preparation of a logical Bell state between two $[[4,1,2]]$ error-detecting code blocks~\cite{andersen2020, marques2022}.

\subsection{Logical Bell-state preparation with the \\$[[4,1,2]]$ code}\label{sec:412_bell_state}
\begin{figure}[t!]
    \centering
    \includegraphics[scale=1]{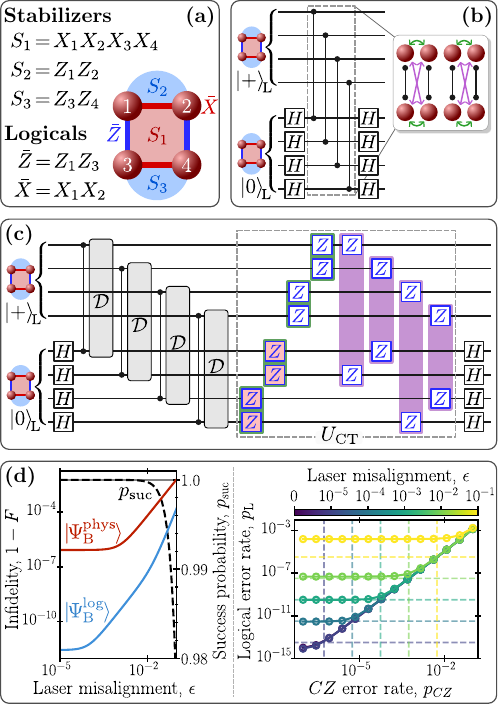}
    \caption{
        Fault-tolerant logical Bell-state preparation between two $[[4,1,2]]$ code blocks in the presence of crosstalk errors.
        \textbf{(a)}~Stabilizer generators and logical operators of the $[[4,1,2]]$ code.
        \textbf{(b)}~FT preparation of a logical Bell state between two code blocks using a transversal \CZ gate.
        As illustrated in the inset, the transversal operation can be implemented in parallel, while imperfect GC due to wave-vector misalignment generates residual intra- and inter-block couplings.
        \textbf{(c)}~Simulated Bell-state preparation circuit including depolarizing \CZ errors and correlated \ZZ errors induced by wave-vector misalignment.
        Of the eight correlated error events, six are either stabilizers or detected by post-selection, whereas the two highlighted in red induce an undetectable logical $X$ error on the second code block.
        \textbf{(d)}~Performance of the Bell-state preparation in the presence of wave-vector misalignment.
        In the left panel, the infidelities of the post-selected logical Bell state (blue) and the corresponding physical Bell state (red) are shown together with the post-selection success probability $p_\mathrm{suc}$ (black dashed).
        Post-selection keeps the logical infidelity below the physical infidelity by detecting the leading-order intrinsic gate errors and a subset of the correlated crosstalk errors.
        In the right panel, the logical error rate $p_\mathrm{L}$ is plotted versus the depolarizing \CZ error rate $p_{\text{\CZ}}$ for different misalignments $\epsilon$.
        Its characteristic quadratic scaling with $p_{\text{\CZ}}$ crosses over to a misalignment-dependent error floor once the correlated \ZZ errors dominate.}
    \label{fig:412_code}
\end{figure}

The $[[4,1,2]]$ error-detecting code, which can be regarded as a minimal distance-two surface-code instance, encodes one logical qubit into four physical qubits as illustrated in Fig.~\ref{fig:412_code}(a).
Its code space is the joint $+1$ eigenspace of the three stabilizers 
\begin{equation}
    S_1 = X_1X_2X_3X_4,\quad
    S_2=Z_1Z_2,\quad
    S_3=Z_3Z_4.
\end{equation}
Any single-qubit Pauli error maps a code state onto a subspace with a nontrivial stabilizer syndrome and can therefore be detected. 
However, the syndrome does not uniquely identify the underlying error, since single-qubit errors that differ by a logical operator can produce the same syndrome.
Consequently, the code can detect, but not reliably correct, an arbitrary single-qubit error.
The code space is spanned by the logical states
\begin{equation}
\begin{aligned}
    \ket{0}_\mathrm{L} &= \frac{1}{\sqrt{2}}\big(\ket{0000}+\ket{1111}\big),\\
    \ket{1}_\mathrm{L} &= \frac{1}{\sqrt{2}}\big(\ket{0011}+\ket{1100}\big)
\end{aligned}
\end{equation}
with logical operators $\bar{Z}=Z_1Z_3$ and $\bar{X}=X_1X_2$.
The logical states can be prepared fault-tolerantly, such that a single physical error occurring during their preparation cannot cause an undetected logical error~\cite{linke2017, zhang2024}.

We consider the preparation of the logical Bell state between two instances of the $[[4,1,2]]$ code
\begin{equation}
    \ket{\Psi_\mathrm{B}}=\frac{1}{\sqrt{2}}\big(\ket{00}_\mathrm{L}+\ket{11}_\mathrm{L}\big)
\end{equation}
using the circuit shown in Fig.~\ref{fig:412_code}(b), which is FT since its transversal structure ensures that any single gate error affects at most one qubit per code block and therefore remains detectable.
The transversal structure of this circuit maps straightforwardly onto the double-well architecture.
Each logical code block is assigned to one ion well containing $n=4$ physical qubits, while the four physical \CZ operations are implemented simultaneously by the inter-well transversal gate.
Thus, the logical Bell-state preparation provides a direct application of the proposed gate within an FT protocol.

\subsection{Crosstalk-induced logical errors}\label{sec:crosstalk_logical_errors}
Misalignment in the wave vector introduces additional inter- and intra-chain \ZZ couplings during the transversal gate, as illustrated in the inset of Fig.~\ref{fig:412_code}(b).
The effect of these errors depends crucially on whether the residual interaction acts within or between the two code blocks, as illustrated in Fig.~\ref{fig:412_code}(c).
The four inter-chain \ZZ errors (violet) act as weight-one errors on each logical code block and therefore remain detectable by the $[[4,1,2]]$ code.
In contrast, the intra-chain \ZZ errors (green) act as weight-two errors within a single code block and are not necessarily detectable.
Those acting on the first logical qubit correspond to stabilizers and hence leave the encoded state unchanged.
Thus, six of the eight possible single crosstalk-error events do not lead to an undetectable logical error.
The remaining two intra-chain errors, highlighted in red in Fig.~\ref{fig:412_code}(c), act on the second logical qubit and propagate through the final transversal Hadamard into logical $X$ errors.
They therefore remain undetected while directly changing the encoded state, providing the relevant mechanism by which wave-vector misalignment can compromise the FT Bell-state preparation.

To quantify the general impact of the misalignment-induced crosstalk on the resulting Bell-state fidelity, we simulate the Bell-state preparation circuit in Fig.~\ref{fig:412_code}(b) by replacing the transversal gate by the unitary evolution under the effective Hamiltonian~\eqref{eqn:effective_hamiltonian} while logical state preparation and single-qubit operations are assumed to be ideal.
We use $\omega_x=\freq{2.0}$, $\omega_y=\freq{2.1}$, and $\omega_z=\freq{0.613}$, while all remaining trap parameters are chosen as before.
We restrict the analysis to detuning scenario~3 and use the decoupling scheme including GC.
Wave-vector misalignment is included through a finite value of $\epsilon$, such that the residual crosstalk couplings are generated directly by the corresponding imperfectly canceled interaction components in the Hamiltonian.
In addition, the nonuniform interaction strengths between opposing ion pairs give rise to the intrinsic transversal-gate infidelity discussed above, which is approximately $10^{-6}$ for the parameters considered here.

After the Bell-state preparation circuit, we post-select on the joint code space of the two logical qubits by projecting the final state $\ket{\psi}$ onto this subspace.
The post-selected state is then given by
\begin{equation}
    \ket{\psi_\mathrm{post}}
    =
    \frac{P_\mathrm{code}\ket{\psi}}
    {\sqrt{p_\mathrm{suc}}},\quad p_\mathrm{suc}=\bra{\psi}P_\mathrm{code}\ket{\psi},
\end{equation}
where $p_\mathrm{suc}$ is the success probability and the projector onto the joint code space reads
\begin{equation}
    P_\mathrm{code}
    =
    \frac{1}{64}
    \prod_{i=1}^{3}(\mathds{I}+S_i^{(1)})(\mathds{I}+S_i^{(2)}).
\end{equation}
Here, $S_i^{(1,2)}$ denotes the $i$th stabilizer of the first and second code block, respectively.
The fidelity of the resulting logical Bell state is evaluated as
\begin{equation}\label{eqn:412_fidelity}
    F=\big|\!\braket{\Psi_\mathrm{B}|\psi_\mathrm{post}}\!\big|^2=\frac{\big|\!\braket{\Psi_\mathrm{B}|\psi}\!\big|^2}{p_\mathrm{suc}}.
\end{equation}

The left panel of Fig.~\ref{fig:412_code}(d) shows the resulting Bell-state infidelity as a function of the wave-vector misalignment $\epsilon$, considering only the intrinsic transversal-gate error and the coherent crosstalk induced by wave-vector misalignment.
For comparison, we show the infidelity of one of the physical Bell pairs generated by the circuit together with the infidelity of the logical Bell state given by Eq.~\eqref{eqn:412_fidelity}.
The corresponding post-selection success probability $p_\mathrm{suc}$ is depicted as well.
In the absence of significant crosstalk, the physical Bell-state infidelity is limited by the intrinsic transversal-gate error and therefore approaches a floor of approximately $10^{-6}$.
The logical Bell-state infidelity remains substantially below the physical one, since the leading-order intrinsic error contributions and four of the eight correlated crosstalk-error events are detected by the code, while two further crosstalk-error events act trivially as stabilizers.
For sufficiently small $\epsilon$, simultaneous crosstalk-error events are strongly suppressed and the misalignment-induced contribution is dominated by individual correlated \ZZ errors.
Since their error probability scales as $J_{\parallel/\perp}^2\propto\epsilon^2$, the logical Bell-state infidelity likewise shows a quadratic dependence on the wave-vector misalignment.
For vanishing misalignment, the remaining logical infidelity is instead set by events containing two intrinsic gate errors.
Since individual intrinsic errors are detected, these contributions occur only at second order, resulting in a logical infidelity floor of approximately $10^{-12}$.

These results demonstrate that the $[[4,1,2]]$ code improves the Bell-state fidelity in the presence of both intrinsic gate errors and misalignment-induced crosstalk.
They do not, however, reveal how crosstalk affects the FT properties of the circuit.
To investigate this, we consider local errors of the physical \CZ gates and crosstalk errors separately using the error model illustrated in Fig.~\ref{fig:412_code}(c).
We model the local error of each physical \CZ operation by an independent two-qubit depolarizing channel with error probability $p_{\text{\CZ}}$,
\begin{equation}
    \mathcal{D}(\rho)=(1-p_{\text{\CZ}})\rho+\frac{p_{\text{\CZ}}}{15}\hspace{-5pt}\sum_{P\in\mathcal{P}_2\setminus\{I\}}\hspace{-10pt}P\rho P,
\end{equation}
where $\mathcal{P}_2=\{I,X,Y,Z\}^{\otimes 2}$ denotes the two-qubit Pauli group. 
Thus, with probability $p_{\text{\CZ}}$, one of the 15 nontrivial two-qubit Pauli errors is applied uniformly at random. 
This effective error model aims at capturing the intrinsic gate error as well as other local error sources, such as entangling pulse-area fluctuations and decoherence. 
Under the depolarizing error model alone, the Bell-state preparation circuit is FT, such that any single gate error remains detectable, including correlated Pauli errors within an entangled qubit pair.
Misalignment-induced crosstalk, however, is not captured by this model, as it arises from residual \ZZ couplings between qubit pairs beyond the intended transversal interactions.
In our simulations, we model these residual couplings as coherent dynamics,
\begin{equation}
    U_\mathrm{CT}=e^{-\ci H_\mathrm{CT}\tau},\quad
    H_\mathrm{CT}=\sum_{i=1}^{N}\!\!\sum_{\substack{j\neq i\\(i,j)\notin\mathcal{T}}}\!\!\!\widetilde{J}_{ij}^{\mathrm{eff}} Z_i Z_j,
\end{equation}
where $\mathcal{T}=\{(i,i+n),(i+n,i)\}_{i=1}^{n}$ denotes the set of intended transversal couplings.
As discussed above, only the two crosstalk errors highlighted in red in Fig.~\ref{fig:412_code}(c) directly induce undetectable logical $X$ errors and can therefore compromise fault tolerance.
Their impact ultimately depends on the strength of the crosstalk relative to the depolarizing error rate.
Hence, we vary $p_\text{\CZ}$ for fixed laser misalignment $\epsilon$ and calculate the resulting logical error rate $p_\mathrm{L}=1-F$, as shown in the right plot in Fig.~\ref{fig:412_code}(d).
The curves are well described by
\begin{equation}
    p_\mathrm{L}(p_\text{\CZ}, \epsilon) \approx A p_\text{\CZ}^2 + p_\mathrm{floor}(\epsilon),
\end{equation}
where the quadratic scaling with $p_\text{\CZ}$ demonstrates fault tolerance against the depolarizing \CZ-gate errors, while the crosstalk noise gives rise to an error floor $p_\mathrm{floor}(\epsilon)$.
For sufficiently small $\epsilon$, this error floor is predominantly determined by the two intra-chain crosstalk error terms that directly induce a logical $X$ error on the second qubit and is approximately described by
\begin{equation}\label{eqn:error_floor}
    p_\mathrm{floor}(\epsilon) = \sin^2(4J_{\parallel}\tau) \approx \frac{\pi^2\delta^2\epsilon^2}{(\omega_\mathrm{COM}+\omega_\mathrm{STR})^2},
\end{equation}
where we assumed uniform mode amplitudes, $4|J_{\parallel}|\tau\ll1$, and $|\epsilon|\ll1$.
The resulting error floors are depicted by the horizontal dashed lines in Fig.~\ref{fig:412_code}(d) and accurately reproduce the numerical results for $p_\text{\CZ}\!\to0$ in most cases.
For larger misalignments, higher-order contributions become relevant, dominated by pairs of inter-chain crosstalk errors that together induce a logical $X$ error, leading to deviations from the predicted error floor.
The apparent error floor for $\epsilon=0$ and the deviation from the analytical prediction for $\epsilon=10^{-5}$ are numerical artifacts caused by the accumulation of rounding errors when constructing the decoupled time-evolution operator through successive matrix multiplications.

The transition from the regime dominated by fault-tolerantly suppressed \CZ-gate errors to the crosstalk-induced error floor occurs around $p_\text{\CZ}\sim4|J_{\parallel}|\tau/\sqrt{A}$.
The approximate transition between these two regimes is indicated by the vertical dashed lines in Fig.~\ref{fig:412_code}(d).
Around these values, the dominant contribution to the logical error gradually shifts from the quadratically suppressed \CZ-gate errors to the FT-breaking crosstalk errors, which ultimately set the error floor for sufficiently small $p_\text{\CZ}$.

\section{Conclusion}\label{sec:conclusion}
In summary, we have theoretically proposed a parallel implementation of a transversal entangling gate for a quantum spring array architecture~\cite{valentini2025}.
By incorporating geometric cancellation based on the intrinsic spatial dependence of the qubit--qubit interactions in the dynamical decoupling scheme, the required transversal coupling pattern can be engineered efficiently.
Compared with a sequential implementation, the proposed scheme substantially reduces the required Rabi frequency and, for many system sizes, also shortens the gate duration.
Moreover, since no additional phase-space trajectory shaping is required, the implementation remains comparatively simple as the system size increases, with the gate time and Rabi-frequency requirements scaling at most linearly with the number of ions.

Among the experimentally relevant error sources considered here, qubit dephasing constitutes the dominant limitation within the investigated parameter regime and has also been identified as a major error source in experimental double-well entangling operations~\cite{valentini2025}.
Since dephasing errors depend on both the gate duration and the number of qubits, they become increasingly relevant when scaling the gate to larger systems.
The interaction-engineering pulse sequences can partially refocus slowly varying magnetic-field fluctuations and thereby increase the effective coherence times of some ions~\cite{uhrig2007, biercuk2009, wang2021, weber2024}, although this protection arises only as a by-product of the DD scheme.
A systematic characterization of these refocusing properties may enable the decoupling sequences to be optimized simultaneously for interaction engineering and enhanced protection against dephasing.
High-fidelity implementations may additionally benefit from other coherence-preserving techniques or shorter gates, for example by exploiting modes with larger frequency splittings or faster detuning configurations.

Beyond the error sources considered explicitly above, motional heating represents an additional experimental consideration for the proposed gate implementation~\cite{valentini2025, sutherland2022, brown2021}.
While sequential implementations may allow recooling between individual two-qubit gates, intermediate recooling would interrupt the coherent dynamics of the parallel gate considered here.
Experiments on double-well trapped-ion systems have shown substantially lower heating rates for the STR mode than for the COM mode, as spatially homogeneous electric-field fluctuations couple only weakly to the out-of-phase STR motion~\cite{valentini2025}.
This suggests that detuning scenario~1, which couples more strongly to the STR mode and also yields shorter gate durations than scenario~3, may be favorable in the presence of heating, although a quantitative comparison requires an explicit heating model.
Heating can further be mitigated by suppressing electric-field and RF noise~\cite{valentini2025}, as well as through improved trap surfaces and cryogenic operation~\cite{brown2021, mcconnell2015}.

We further assume instantaneous and perfect decoupling $X$-pulses.
During more realistic finite pulses the active qubit--qubit interactions lead to additional coherent phase errors, while pulse-area and control errors can cause small over- or under-rotations and incomplete cancellation of unwanted interactions.
These residual couplings are particularly relevant for FT circuits, as they can generate crosstalk beyond the intended transversal pairs and thereby compromise fault tolerance, analogous to the mechanism discussed in Sec.~\ref{sec:misaligned_wave_vector}.
Coherent errors may be reduced through randomized or appropriately designed pulse sequences that prevent their coherent accumulation~\cite{genov2017, santos2006, yi2026}.
In particular, errors arising from finite pulses can be addressed by explicitly accounting for the interaction dynamics during the pulses or be mitigated by switching off the state-dependent force during the pulses~\cite{zhou2020, manovitz2017}.
More generally, even with ideal decoupling pulses, the all-to-all interactions present during the gate can also cause local errors that do not commute with the Ising interaction to propagate to multiple qubits, thereby generating correlated errors beyond the circuit-level depolarizing noise model considered here.
A more complete time-resolved analysis of the gate dynamics, including finite-duration and imperfect control pulses as well as error propagation during the interaction-engineering sequence, therefore provides a natural direction for further investigation.

Trap-frequency noise constitutes an additional error source, as motional-mode frequency fluctuations perturb the laser detunings and thereby both the effective interactions and phase-space dynamics~\cite{sutherland2022}.
This can lead to phase errors and incomplete trajectory closure, particularly when many phase-space cycles are performed.
Finally, photon scattering and residual AC Stark shifts, which are not included in the effective gate model, constitute additional experimental error sources~\cite{ozeri2007,balance2016}.
In the large-detuning regime considered here, however, photon scattering is strongly suppressed, while AC Stark shifts can in principle be compensated experimentally.

Controlling these additional error sources, and correlated crosstalk errors in particular, will be important for preserving the FT properties of the proposed parallel implementation.
Since transversal entangling gates are a key building block of many QEC protocols, their efficient physical realization across multiple qubit pairs is essential for scalable FT quantum computing.
The scheme introduced here provides a route toward realizing such operations natively and in parallel in trapped-ion systems, avoiding their decomposition into sequences of individual two-qubit gates.

\section*{Data Availability}
All data presented in this manuscript are publicly available on Zenodo~\cite{Zenodo2026}.

\section*{Author contributions}
K.B. developed the gate implementation, performed the analytical calculations and numerical simulations, analyzed the results, prepared the figures, and wrote the manuscript.
A.K. performed the numerical simulations of the spectator-mode effects, while K.B. derived the corresponding analytical approximation. A.B. and M.M. supervised the project.
All authors discussed the results and reviewed and approved the final manuscript.

\section*{Acknowledgments}
K.B. thanks Friederike Butt for helpful discussions.
All authors gratefully acknowledge support by the European Union’s Horizon Europe research and innovation program under Grant Agreement Number 101114305 (“MILLENION-SGA1” EU Project).
K.B. and M.M. acknowledge funding from the Deutsche Forschungsgemeinschaft (DFG, German Research Foundation) through Research Unit FOR 5413/1, Project No.~465199066, Priority Programme SPP 2514, Project No.~563402549, and Germany's Excellence Strategy through the Cluster of Excellence Matter and Light for Quantum Computing (ML4Q), EXC 2004/2, Project No.~390534769. They also acknowledge support from the Federal Ministry of Research, Technology and Space of Germany (BMFTR) through the projects NeuQuant and SQale, and from the Entangled Logical Qubits (ELQ) program, managed by the Intelligence Advanced Research Projects Activity (IARPA).
This research is also part of the Munich Quantum Valley (K-8), which is supported by the Bavarian state government with funds from the Hightech Agenda Bayern Plus.
The authors gratefully acknowledge computing time provided at the NHR Center NHR4CES at RWTH Aachen University (Project No. p0020074), which is funded by the Federal Ministry of Education and Research and the state governments participating on the basis of the resolutions of the GWK for national high-performance computing at universities.
A.K. and A.B. acknowledge support from ID2024-161474NB-I00 (MCIU/AEI/FEDER,UE) and  QUITEMAD-CM TEC-2024/COM-84, from the Grant IFT Centro de Excelencia Severo Ochoa CEX2020-001007-S funded by MCIN/AEI/10.13039/501100011033, and from the CSIC Research Platform on Quantum Technologies PTI-001.

\newpage

\bibliographystyle{apsrev4-2}
\bibliography{references}

\clearpage
\onecolumngrid

\appendix

\section{Derivation of the State-Dependent Dipole Hamiltonian}\label{app:dipole-Hamiltonian-derivation}
In this appendix, we derive the state-dependent dipole Hamiltonian in Eq.~\eqref{eqn:dipole_hamiltonian}.
For notational simplicity, we omit the ion index $i$ from the internal states throughout the following derivation.
We consider the three-level system formed by the two qubit states $\ket{0}$ and $\ket{1}$ and an auxiliary excited state $\ket{r}$.
The dynamics of the internal electronic states are described by the master equation
\begin{equation}
    \frac{\mathrm{d}\rho}{\mathrm{d}t}=-\ci\big[H_0+V,\rho\big]+\mathcal{D}(\rho),
\end{equation}
where the coherent part of the evolution is given by 
\begin{equation}
    H_0=\sum_{q=0,1,r}\epsilon_q \ket{q}\!\!\bra{q}, \quad V=\frac{1}{2}\sum_{l=1,2}\sum_{q=0,1}\Omega_{l,q}\ket{r}\!\!\bra{q}e^{-\ci\omega_l t}+\mathrm{H.c.}.
\end{equation}
Here, $\Omega_{l,q}$ denotes the Rabi frequency of laser $l$ coupling the qubit state $\ket{q}$ to the excited state $\ket{r}$, $\omega_l$ is the laser frequency, and $\epsilon_q$ is the energy of state $\ket{q}$.
The spontaneous decay of the excited state back to the qubit manifold is described by the Lindblad dissipator
\begin{equation}
    \mathcal{D}(\rho)=\sum_{n=0,1}\left(L_n \rho L_n^\dagger-\frac{1}{2}L_n^\dagger L_n\rho-\frac{1}{2}\rho L_n^\dagger L_n\right),
\end{equation}
where we, for simplicity, assume that the excited state decays into each qubit state with rate $\Gamma_r$, such that the total decay rate is~$2\Gamma_r$. 
The corresponding jump operators are $L_0=\sqrt{\Gamma_r}\ket{0}\!\bra{r}$ and $L_1=\sqrt{\Gamma_r}\ket{1}\!\bra{r}$.
For each laser-induced transition, we define the detunings $\delta_{l,q} = \epsilon_r-\epsilon_q-\omega_l$ and assume that these are much larger than the Rabi frequencies and the decay rate, $|\delta_{l,q}|\gg|\Omega_{l,q}|,\Gamma_r$, such that we can adiabatically eliminate the excited state from the dynamics.

Following the effective operator formalism for open quantum systems introduced in Ref.~\cite{reiter2012}, we decompose the Hamiltonian $H=H_0 + V$ into a ground- and excited-state subspace and write
\begin{equation} 
    H = H_\mathrm{g} + H_\mathrm{e} + V_+ + V_-,
\end{equation}
where $H_{g,e}$ describe the interactions inside the ground and excited subspace and $V_+$ and $V_-=V_+^\dagger$ couple these two subspaces.
Adiabatic elimination of the excited subspace generally leads to an effective master equation within the ground-state subspace,
\begin{equation}
    \frac{\mathrm{d}\rho}{\mathrm{d}t} = -\ci\big[H^\mathrm{eff},\rho\big] + \sum_{n=0,1} \left(L_n^\mathrm{eff}\rho(L_n^\mathrm{eff})^\dagger-\frac{1}{2}(L_n^\mathrm{eff})^\dagger L_n^\mathrm{eff}\rho - \frac{1}{2}\rho (L_n^\mathrm{eff})^\dagger L_n^\mathrm{eff} \right)
\end{equation}
with the effective Hamiltonian and Lindblad operators
\begin{align}
    H^{\mathrm{eff}}&=-\frac{1}{2}\left[V_{-}\sum_{l,q}\left(H_{\mathrm{NH}}^{(l, q)} \right)^{-1}V_{+}^{(l,q)}+ \mathrm{H.c.}\right]+ H_\mathrm{g},\\
    L^{\mathrm{eff}}_n&=L_n\sum_{l,q}\left( H_{\mathrm{NH}}^{(l,q)} \right)^{-1}V_{+}^{(l,q)}, \label{eqn:L_eff}
\end{align}
and $V_\pm = \sum_{l,q}V^{(l,q)}_\pm$, where $q$ runs over all the ground states with energy $\epsilon_q$ and $l$ over the excitation fields with frequencies $\omega_l$.
The field- and state-dependent propagators are given by
\begin{equation}
    \left(H_{\mathrm{NH}}^{(l,q)} \right)^{-1}\equiv\left(H_{\mathrm{NH}} - \epsilon_q - \omega_l \right)^{-1}
\end{equation}
with the non-Hermitian Hamiltonian $H_\mathrm{NH} = H_\mathrm{e} - \frac{\ci}{2}\sum_n L_n^\dagger L_n$.

Applying this formalism to the considered three-level system with $V_+^{(l,q)} = \frac{1}{2}\Omega_{l,q}\ket{r}\!\!\bra{q}e^{-\ci\omega_l t}$, the resulting effective Hamiltonian can be separated into
\begin{equation}
    H^\mathrm{eff} = H_\mathrm{S} + H_\mathrm{R} + H_\mathrm{d},
\end{equation}
where $H_\mathrm{S}$ contains the static AC Stark shifts, $H_\mathrm{R}$ describes two-photon Raman transitions between the qubit states, and $H_\mathrm{d}$ gives rise to the state-dependent dipole force.
The AC Stark contribution is given by
\begin{equation}\label{eqn:Hs}
    H_\mathrm{S} = \sum_{q=0,1} \big(\epsilon_q+\Delta\epsilon_q\big) \ket{q}\!\!\bra{q} \quad \text{with} \quad \Delta\epsilon_q = -\frac{1}{4} \sum_{l=1,2} \frac{\abs{\Omega_{l,q}}^2\delta_{l,q} }{\delta_{l,q}^2+\Gamma_r^2}
\end{equation}
and the Hamiltonian describing the two-photon Raman transitions and the corresponding effective Rabi frequencies read
\begin{equation}
    H_\mathrm{R} = \frac{1}{2} \sum_{l,l'} \Omega_{l,l'}^\mathrm{R} \ket{0}\!\!\bra{1} e^{\ci(\omega_l-\omega_{l'})t} + \mathrm{H.c.}\quad \text{with} \quad \Omega_{l,l'}^\mathrm{R} = -\frac{1}{4} \frac{\Omega_{l,0}^*\Omega_{l',1}\big(\delta_{l,0}+\delta_{l',1}\big) }{\big(\delta_{l,0}+\ci\Gamma_r\big)\big(\delta_{l',1}-\ci\Gamma_r\big) }.
\end{equation}
Of particular interest for the present work is the remaining contribution, which is the state-dependent dipole interaction given by
\begin{equation}
    H_\mathrm{d} = \frac{1}{2} \sum_{q=0,1} \Omega_q^\mathrm{d} \ket{q}\!\!\bra{q} e^{\ci(\omega_1-\omega_2)t} + \mathrm{H.c.} \quad \text{with} \quad \Omega_q^\mathrm{d} = -\frac{1}{4} \frac{\Omega_{1,q}^*\Omega_{2,q}\big(\delta_{1,q}+\delta_{2,q}\big) }{\big(\delta_{1,q}+\ci\Gamma_r\big)\big(\delta_{2,q}-\ci\Gamma_r\big)}.
\end{equation}
Using $Z = \ket{0}\!\bra{0} - \ket{1}\!\bra{1}$ and the laser beatnote $\omega_\mathrm{L}=\omega_1-\omega_2$, the dipole contribution can be decomposed as
\begin{equation}
    H_\mathrm{d} = \frac{1}{2} \widetilde{\Omega}_\mathrm{L} \mathds{I}\, e^{-\ci\omega_\mathrm{L}t} + \frac{1}{2} \Omega_\mathrm{L} Z e^{-\ci\omega_\mathrm{L}t} + \mathrm{H.c.},
\end{equation}
where $\widetilde{\Omega}_\mathrm{L}^* = \frac{1}{2} \big(\Omega_0^{\mathrm{d}}+\Omega_1^{\mathrm{d}} \big), \, \Omega_\mathrm{L}^* = \frac{1}{2} \big(\Omega_0^\mathrm{d}-\Omega_1^\mathrm{d} \big)$.
While the first term is independent of the qubit state, the second term produces a state-dependent dipole force.
Based on the laser beatnote $\omega_\mathrm{L}$, the dynamics generated by the effective laser--ion interaction can be selected.
For $\omega_\mathrm{L}\approx\omega_0$, with $\omega_0=\epsilon_1-\epsilon_0$, the two-photon Raman term $H_\mathrm{R}$ becomes resonant and drives transitions within the qubit subspace.
However, in the regime considered throughout this work, $\omega_\mathrm{L}\ll\omega_0$, these Raman transitions are far off-resonant and can be neglected, leaving the state-dependent dipole interaction $H_\mathrm{d}$.

After deriving the coherent part of the effective dynamics, we compute the jump operators in Eq.~\eqref{eqn:L_eff} to obtain the effective dissipative dynamics, which read
\begin{align}
L_0^{\mathrm{eff}} &=\frac{1}{2}\sqrt{\Gamma_r}\left(\frac{\Omega_{1,0} e^{-\ci\omega_1 t}}{\delta_{1,0}-\ci\Gamma_r}+\frac{\Omega_{2,0} e^{-\ci\omega_2 t}}{\delta_{2,0}-\ci\Gamma_r}\right)\lvert 0\rangle\!\langle 0\rvert+\frac{1}{2}\sqrt{\Gamma_r}\left(\frac{\Omega_{1,1} e^{-\ci\omega_1 t}}{\delta_{1,1}-\ci\Gamma_r}+\frac{\Omega_{2,1} e^{-\ci\omega_2 t}}{\delta_{2,1}-\ci\Gamma_r}\right)\lvert 0\rangle\!\langle 1\rvert,\\
L_1^{\mathrm{eff}} &=\frac{1}{2}\sqrt{\Gamma_r}\left(\frac{\Omega_{1,1} e^{-\ci\omega_1 t}}{\delta_{1,1}-\ci\Gamma_r}+\frac{\Omega_{2,1} e^{-\ci\omega_2 t}}{\delta_{2,1}-\ci\Gamma_r}\right)\lvert 1\rangle\!\langle 1\rvert + \frac{1}{2}\sqrt{\Gamma_r}\left(\frac{\Omega_{1,0} e^{-\ci\omega_1 t}}{\delta_{1,0}-\ci\Gamma_r}+\frac{\Omega_{2,0} e^{-\ci\omega_2 t}}{\delta_{2,0}-\ci\Gamma_r}\right)\lvert 1\rangle\!\langle 0\rvert.
\end{align}
The first term of each of the jump operators corresponds to the Rayleigh photon scattering, under which the internal ion state is preserved.
The second term in each operator describes the Raman scattering, which changes the qubit state under emission of a photon.

Up to this point, we assumed that the ions are fixed in space and neglected their motional degrees of freedom.
To account for the coupling of the laser fields to the motion of the ions, the spatial phase of the laser fields is included according to $\Omega_{l,q} \rightarrow \Omega_{l,q}e^{\ci\mathbf{k}_l\cdot\mathbf{r}}$.
Defining the effective wave vector $\mathbf{k}_\mathrm{L} = \mathbf{k}_1-\mathbf{k}_2$, the dipole Hamiltonian becomes
\begin{equation}\label{eqn:eff_Hd}
    H_\mathrm{d} = \frac{1}{2} \widetilde{\Omega}_\mathrm{L}  \mathds{I}\, e^{\ci\mathbf{k}_\mathrm{L}\cdot\mathbf{r}} e^{-\ci\omega_\mathrm{L}t} + \frac{1}{2} \Omega_\mathrm{L} Z e^{\ci\mathbf{k}_\mathrm{L}\cdot\mathbf{r}} e^{-\ci\omega_\mathrm{L}t} + \mathrm{H.c.}.
\end{equation}
The state-independent contribution in Eq.~\eqref{eqn:eff_Hd} does not affect the internal qubit dynamics but couples only to the vibrational degrees of freedom. 
Together with the static AC Stark shifts contained in Eq.~\eqref{eqn:Hs}, these contributions are assumed to be compensated experimentally by an appropriate choice of the laser parameters.

Finally, we consider the large-detuning regime, $\Gamma_r \ll \abs{\delta_{l,q}}$, with $\delta_{l,q}\approx\Delta$, such that
$\Omega_q^\mathrm{d}\approx - \Omega_{1,q}^*\Omega_{2,q}/(2\Delta)$ and the differential two-photon Rabi frequency can be written as
\begin{equation} 
    \Omega_\mathrm{L} = \frac{\Omega_{1,1}\Omega_{2,1}^* - \Omega_{1,0}\Omega_{2,0}^* }{4\Delta}.
\end{equation}
Writing the position of ion $i$ as $\mathbf{r}_i = \mathbf{r}_i^{(0)} + \delta\mathbf{r}_i$ and summing over all ions then yields
\begin{equation}
    H_\mathrm{d} = \frac{1}{2} \sum_i \Omega_\mathrm{L} e^{\ci\mathbf{k}_\mathrm{L}\cdot\mathbf{r}_i^{(0)}} e^{\ci\big(\mathbf{k}_\mathrm{L}\cdot\delta\mathbf{r}_i - \omega_\mathrm{L}t \big)} Z_i + \mathrm{H.c.},
\end{equation}
which corresponds to Eq.~\eqref{eqn:dipole_hamiltonian} and describes a time- and state-dependent AC Stark shift induced by the interference of the two laser beams.
The effective jump operators within the considered regime read
\begin{align}
L_0^{\mathrm{eff}} &=\frac{\sqrt{\Gamma_r}}{2\Delta}\left(\Omega_{1,0} e^{\ci(\mathbf{k}_1\cdot \mathbf{r}-\omega_1 t)}+\Omega_{2,0} e^{\ci(\mathbf{k}_2\cdot \mathbf{r}-\omega_2 t)}\right)\lvert 0\rangle\!\langle 0\rvert+\frac{\sqrt{\Gamma_r}}{2\Delta}\left(\Omega_{1,1} e^{\ci(\mathbf{k}_1\cdot \mathbf{r}-\omega_1 t)}+\Omega_{2,1} e^{\ci(\mathbf{k}_2\cdot \mathbf{r}-\omega_2 t)}\right)\lvert 0\rangle\!\langle 1\rvert,\\
L_1^{\mathrm{eff}} &=\frac{\sqrt{\Gamma_r}}{2\Delta}\left(\Omega_{1,1} e^{\ci(\mathbf{k}_1\cdot \mathbf{r}-\omega_1 t)}+ \Omega_{2,1} e^{\ci(\mathbf{k}_2\cdot \mathbf{r}-\omega_2 t)}\right)\lvert 1\rangle\!\langle 1\rvert + \frac{\sqrt{\Gamma_r}}{2\Delta}\left(\Omega_{1,0} e^{\ci(\mathbf{k}_1\cdot \mathbf{r}-\omega_1 t)}+\Omega_{2,0} e^{\ci(\mathbf{k}_2\cdot \mathbf{r}-\omega_2 t)}\right)\lvert 1\rangle\!\langle 0\rvert,
\end{align}
where it becomes clear that the effective scattering rate scales with $\Gamma_r (|\Omega_{l,q}|/\Delta)^2$ and is therefore strongly suppressed in the large-detuning regime considered here.

\section{Magnus expansion of the dipole Hamiltonian}\label{app:magnus-expansion}
The state-dependent force described by the dipole Hamiltonian in Eq.~\eqref{eqn:dipole_Hamiltonian_LDA} gives rise to both a state-dependent displacement of the motional modes and an effective qubit--qubit interaction.
To characterize both contributions, we calculate the corresponding time-evolution operator using a Magnus expansion.
This allows us to derive the effective interaction Hamiltonian and identify the condition under which the phase-space trajectories close and the qubit and vibrational degrees of freedom disentangle at the end of the gate.

Starting from the dipole Hamiltonian in Eq.~\eqref{eqn:dipole_Hamiltonian_LDA},
\begin{equation}
    H_\mathrm{d}(t)=\sum_{i,m}(g_{i,m}e^{\ci\delta_m t}b_m^\dagger+g_{i,m}^*e^{-\ci\delta_m t}b_m)Z_i,
\end{equation}
where $m\in\{\mathrm{COM},\mathrm{STR}\}$, the commutator of the Hamiltonian at two different times is given by
\begin{equation}
    [H_\mathrm{d}(t_1),H_\mathrm{d}(t_2)]=\sum_{i,j,m}\Big(g_{i,m}^*g_{j,m}e^{-\ci\delta_m(t_1-t_2)}-g_{j,m}^*g_{i,m}e^{\ci\delta_m(t_1-t_2)}\Big)Z_iZ_j.
\end{equation}
Since this expression contains no bosonic mode operators and $[Z_i, Z_j]=0$, it commutes with $H_\mathrm{d}(t)$ at all times, such that all subsequent terms in the Magnus expansion vanish. 
The Magnus expansion therefore terminates after second order, such that the time-evolution operator can be written exactly as $U(t)=\exp(\Omega_1(t)+\Omega_2(t))$, with
\begin{align}\label{eq_magnus_expansion}
    \Omega_1(t)&=\int_0^t \mathrm{d}t_1 \tilde{H}(t_1)= \sum_{i,m}\bigg(\frac{1-e^{\ci\delta_m t}}{\delta_m}g_{i,m}b_m^\dagger-\frac{1-e^{-\ci\delta_m t}}{\delta_m}g_{i,m}^*b_m\bigg)Z_i\\
    \Omega_2(t)&= \frac{1}{2}\int_0^t \mathrm{d}t_1 \int_0^{t_1} \mathrm{d}t_2 \, [\tilde{H}(t_1), \tilde{H}(t_2)] =\ci\sum_{i,j,m}g_{i,m}^*g_{j,m}\left(\frac{t}{\delta_m}-\frac{\sin(\delta_m t)}{\delta_m^2}\right)Z_iZ_j,
\end{align}
where $\tilde{H}=-iH$.
Here, $\Omega_1(t)$ describes a state-dependent displacement of the motional modes in phase space, whereas $\Omega_2(t)$ gives rise to the geometric phase underlying the effective qubit--qubit interaction. 
Since $[\Omega_1(t),\Omega_2(t)]=0$, the propagator factorizes according to $U(t)=\exp(\Omega_1(t))\exp(\Omega_2(t))$.

The motional part of the evolution can thus be expressed as a state-dependent displacement operator,
\begin{equation}
    U_\mathrm{p}(t)=\exp(\Omega_1(t))=\exp\bigg(\sum_{i,m}\Big(\alpha_{i,m}(t)b_m^\dagger-\alpha_{i,m}^*(t)b_m\Big)Z_i\bigg),
\end{equation}
where
\begin{equation}\label{eq_phase_space_displacement}
    \alpha_{i,m}(t)=\frac{1-e^{\ci\delta_m t}}{\delta_m}g_{i,m}
\end{equation}
is the displacement amplitude associated with ion $i$ and mode $m$. 
The corresponding phase-space trajectory is closed whenever the displacement vanishes.
For the qubit and motional degrees of freedom to be disentangled at the end of the gate, this condition must be satisfied for every driven mode, which requires $e^{\ci\delta_m\tau}=1$ and therefore $|\delta_m|\tau=2\pi k_m$, with $k_m\in\mathds{N}$ the number of cycles mode $m$ performs in phase space.
Hence, a mode driven with detuning $\delta_m$ completes one closed phase-space trajectory after a time
\begin{equation}
    \tau_m=\frac{2\pi}{|\delta_m|}.
\end{equation}
A common gate duration must consequently be chosen such that it corresponds to an integer number of phase-space cycles for both the COM and STR modes. 

At such times, the state-dependent displacement vanishes and the remaining evolution is governed entirely by the second Magnus term.
Since $\sin(\delta_m\tau)=0$ for all driven modes, $\Omega_2(\tau)$ reduces to
\begin{equation}
    \Omega_2(\tau)=\ci\tau\sum_{i,j,m}\frac{g_{i,m}^*g_{j,m}}{\delta_m}Z_iZ_j.
\end{equation}
Identifying the remaining evolution with $U(\tau)=\exp(-\ci H_\mathrm{eff}\tau)$ and inserting $g_{i,m} = \frac{\ci}{2}\Omega_\mathrm{L}\eta_m^z e^{\ci\mathbf{k}_\mathrm{L}\cdot\mathbf{r}_i^{(0)}}\mathcal{M}_{i,m}^z$ yields the effective Ising interaction
\begin{equation}
    H_\mathrm{eff}=\sum_{i=1}^N\sum_{j\neq i}J_{ij}^\mathrm{eff}Z_iZ_j
\end{equation}
with
\begin{equation}
    J_{ij}^\mathrm{eff}=\chi_{ij}\cos\bigl(\mathbf{k}_\mathrm{L}\cdot\mathbf{r}_{ij}^{(0)}\bigr),\qquad\chi_{ij}=-\sum_m\frac{\Omega_\mathrm{L}^2\eta_m^2}{4\delta_m}\mathcal{M}_{i,m}\mathcal{M}_{j,m}.
\end{equation}
Here, the contributions from $(i,j)$ and $(j,i)$ have been combined, while the terms with $i=j$ have been omitted since they only give rise to a global phase.
This corresponds to the effective Hamiltonian given in Eq.~\eqref{eqn:effective_hamiltonian}.

\section{Infidelity induced by entangling pulse-area fluctuations}\label{app:fluctuation_error}
In this appendix, we derive the gate infidelity resulting from quasi-static fluctuations of the entangling pulse area.
We assume perfect decoupling and equal interaction strengths between opposing ion pairs, i.e., $J_{i,i+n}=J$ for $i=1,\ldots,n$.
Evaluating the time evolution of the transversal gate in Eq.~\eqref{eqn:time_evolution} at the gate time $\tau=\pi/(8|J|)$ yields the target entangling operation
\begin{equation}
    U_\mathrm{T}=\prod_{i=1}^{n}\exp\Big(-\ci\,\mathrm{sign}(J)\frac{\pi}{4}Z_iZ_{i+n}\Big).
\end{equation}

We consider a relative fluctuation of the effective interaction strength, $J\rightarrow J(1+\epsilon_J)$, where $\epsilon_J$ denotes the relative deviation from the target interaction strength.
The gate duration $\tau=\pi/(8|J|)$ is fixed by the target interaction strength defined in Eq.~\eqref{eqn:J_approx} and remains unchanged by the fluctuation.
The modified interaction strength results in the evolution
\begin{equation}
    U(\tau)=\prod_{i=1}^{n}\exp\Big[-\ci\,\mathrm{sign}(J)\frac{\pi}{4}(1+\epsilon_J)Z_iZ_{i+n}\Big]=U_\mathrm{T}\prod_{i=1}^{n}\exp\Big(-\ci\,\mathrm{sign}(J)\frac{\pi}{4}\epsilon_J Z_iZ_{i+n}\Big).
\end{equation}
Thus, the pulse-area fluctuation introduces an additional \ZZ phase $\varphi_J=\pi\epsilon_J/4$ for each pair of opposing ions.
Within this effective description, the additional phase depends only on the relative pulse-area fluctuation~$\epsilon_J$.
The state fidelity with respect to the target state \eqref{eqn:target_state} is given by
\begin{equation}
    F_\mathrm{fluct}=\big|\bra{\Psi_0}U_\mathrm{T}^\dagger U(\tau)\ket{\Psi_0}\big|^2= \Big| \bra{\Psi_0} \prod_{i=1}^{n} \exp\Big( -\ci\,\mathrm{sign}(J)\frac{\pi}{4}\epsilon_J Z_iZ_{i+n} \Big) \ket{\Psi_0} \Big|^2
\end{equation}
with the initial state $\ket{\Psi_0}=\ket{+}^{\otimes 2n}$.
Since this state factorizes into $n$ two-ion product states, $\ket{\Psi_0}=\bigotimes_{i=1}^{n}\ket{++}_{i,i+n}$, and the individual terms act on disjoint ion pairs, the fidelity can be written as
\begin{align}
    F_\mathrm{fluct} &= \prod_{i=1}^{n} \Big| \bra{++} \exp\Big( -\ci\,\mathrm{sign}(J)\frac{\pi}{4}\epsilon_J (Z\otimes Z) \Big) \ket{++} \Big|^2 \\
    &= \prod_{i=1}^{n} \Big| \bra{++} \Big[ \cos\Big(\mathrm{sign}(J)\frac{\pi}{4}\epsilon_J\Big)\mathds{I} - \ci\sin\Big(\mathrm{sign}(J)\frac{\pi}{4}\epsilon_J\Big) (Z\otimes Z) \Big] \ket{++} \Big|^2 \notag\\
    &=\cos^{2n}\Big(\frac{\pi}{4}\epsilon_J\Big),
\end{align}
where we used $\bra{++}Z\otimes Z\ket{++}=0$ and the parity of the cosine.
For small pulse-area fluctuations, $|\epsilon_J|\ll1$, expansion to second order yields
\begin{equation}
    F_\mathrm{fluct}\approx 1-\frac{n\pi^2}{16}\epsilon_J^2.
\end{equation}

Finally, assuming normally distributed fluctuations, $\epsilon_J\sim\mathcal{N}(0,\sigma_J^2)$, we have the expectation value $\mathbb{E}[\epsilon_J^2]=\sigma_J^2$.
The expected gate fidelity therefore becomes
\begin{equation}
    \bar{F}_\mathrm{fluct}=\mathbb{E}\big[F_\mathrm{fluct}\big]\approx 1-\frac{n\pi^2}{16}\sigma_J^2,
\end{equation}
which corresponds to Eq.~\eqref{eqn:fluctuation_fidelity}.
The mean infidelity caused by entangling pulse-area fluctuations is therefore given by $1-\bar{F}_\mathrm{fluct}$.

\section{Coupling to spectator modes}\label{app:spectator_modes}
In this appendix, we derive the analytical expression for the fidelity in the presence of spectator-mode coupling used in Sec.~\ref{sec:spectator_mode_coupling}. 
We first derive the state fidelity for a single interval of free time evolution between consecutive DD pulses and then extend this result to the full DD sequence by accounting for the intermediate $X$-pulses. 
Finally, we expand the fidelity for small phase-space displacements.

For a period of free time evolution between two decoupling pulses, the fidelity can be computed from the time-evolution operator $U(t,t_0)= \exp(\Omega_1(t,t_0)) \exp(\Omega_2(t,t_0))$ as derived via the Magnus expansion in App.~\ref{app:magnus-expansion}.
We assume the initial qubit state $\ket{+}^{\otimes N}=\frac{1}{2^{N/2}} \sum_{\{s\}} \ket{s}$, with $\ket{s}=\ket{s_1,\dots, s_N}$, in terms of the computational basis states with $Z_i\ket{s_i}=s_i\ket{s_i}$, $s_i=\pm1$.
The time-evolution operator including the $N$ axial modes reads
\begin{equation}
U(t,t_0) = \sum_{\{s\}} e^{-\ci\theta_s(t,t_0)} \ket{s}\bra{s} \otimes \prod_{m=1}^{N}D_m(\beta_{ms}),
\end{equation}
where $\theta_s(t,t_0) = \sum_{ij} s_i s_j \left[ (t-t_0)J_{ij} + K_{ij}(t,t_0) \right]$. 
The matrices $J_{ij}$ and $K_{ij}$ describing the qubit--qubit couplings are defined by
\begin{equation}
\begin{aligned}
   J_{ij} & =- \sum_{m=1}^N \frac{F_{i,m} F_{j,m}^* }{\delta_m}\cos(\Phi_{ij}), \\
    K_{ij}(t,t_0) & = \sum_{m=1}^N \frac{F_{i,m} F_{j,m}^* }{\delta_m^2} (\sin(\delta_m (t-t_0) + \Phi_{ij} ) - \sin\Phi_{ij}),
\end{aligned}
\end{equation}
where we introduced $F_{i,m}= \ci \frac{\Omega_\mathrm{L}}{2} \eta_m^z \mathcal{M}_{i,m}^z$ and $\Phi_{ij}=\mathbf{k}_\mathrm{L} \cdot (\mathbf{r}_i^{(0)}-\mathbf{r}_j^{(0)})$.
Note that the intrinsic phase errors discussed in the main text resulting from nonuniform transversal couplings between opposing ions $J_{i, i+n}$ are also included in the following analysis.
The exponential $\exp(\Omega_1(t,t_0))$ couples qubit and vibrational degrees of freedom through the displacement operators $D_m(\beta)=e^{\beta b_m^{\dagger}-\beta^*b_m}$ of mode $m$.
For a computational state $s$, the corresponding displacement is $\beta_{ms}= \sum_{j=1}^N \alpha_{jm}(t,t_0)s_j$ with $\alpha_{j,m}(t,t_0)=(e^{\ci\delta_m t_0}-e^{\ci\delta_m t})g_{j,m}/\delta_m$.
Considering only the COM and STR modes driving the gate, the oscillating term $K_{ij}$ vanishes at the phase-space closure points given by Eq.~\eqref{eqn:gate_time_ps}, and only the constant qubit--qubit coupling $J_{ij}$ remains. 

The target-state fidelity is obtained from the overlap between the ideal target state and the state resulting from the actual time evolution as $F=\bra{\Psi_\mathrm{T}} \mathrm{Tr}_\mathrm{ph}\left[ U(t,t_0)  \rho_0 U^\dagger(t,t_0) \right]\ket{\Psi_\mathrm{T}}$. Here, the initial state is chosen as $\rho_0=(\ket{+}\!\!\bra{+})^{\otimes N} \otimes \rho_\mathrm{th} $, where  $\rho_\mathrm{th}$ is the thermal state of the phonon modes, leading to
\begin{equation}\label{eq_fid_spectator_modes_single_interval}
F = \frac{1}{2^{2N}}\sum_{\{s, r\}}e^{-\ci(\theta_s-\theta_{r})} e^{\ci (\Delta_s - \Delta_r)}\prod_{m=1}^{N}\langle D_m^{\dagger}(\beta_{mr})D_m(\beta_{ms})\rangle_\mathrm{th},
\end{equation}
where $\ket{\Psi_\mathrm{T}}=\exp\left(\ci\frac{\pi}{4}\sum_{i=1}^n Z_i Z_{i+n}\right)\ket{+}^{\otimes N}$ is the target state obtained by applying the ideal transversal \CZ gate.
For detuning scenario~3 considered here, the effective interaction has $J<0$, such that $\mathrm{sign}(J)=-1$.
Moreover, $\langle O \rangle_\mathrm{th}$ denotes the thermal expectation value of the operator $O$ with respect to the thermal state $\rho_\mathrm{th}$.
The phase $\Delta_s=\sum_{i,j=1}^{N} s_i s_j B_{ij}$ describes the phase acquired during the evolution under the ideal gate, where the matrix elements are defined as $B_{i,i+n}=B_{i+n,i}=-\pi/8$ for $i=1,\dots,n$, while all other entries vanish.
Finally, for a thermal state with mean phonon number $\bar{n}_m$, the expectation value of a single-mode displacement is given by \cite{scully1997}
\begin{equation}\label{eq_thermal_exp_value}
\langle D_m(\alpha)\rangle_{\mathrm{th}} = \exp\left(-|\alpha|^2(\bar{n}_m+1/2)\right),
\end{equation}
which allows for a straightforward evaluation of the fidelity
\begin{equation}\label{eqn:fidelity_free_evolution}
F = \frac{1}{2^{2N}} \sum_{\{s, r\}}e^{-\ci(\theta_s-\theta_{r})} e^{\ci (\Delta_s - \Delta_r)}\prod_{m=1}^{N} \exp\left({(-\beta_{mr} \beta_{ms}^* + \beta_{mr}^* \beta_{ms})/2}\right) \exp\left({-|\beta_{ms}- \beta_{mr}|^2(\bar{n}_m+1/2)}\right),
\end{equation}
where we used $D^\dagger(\alpha)=D(-\alpha)$ and $D(\alpha)D(\beta)= e^{(\alpha \beta^* - \alpha^* \beta)/2} D(\alpha + \beta)$.
In this expression, spectator-mode coupling affects the fidelity through additional contributions to the accumulated qubit--qubit phase and through residual spin--motion entanglement associated with unclosed phase-space trajectories.

We now generalize Eq.~\eqref{eq_fid_spectator_modes_single_interval} to account for the decoupling $X$-pulse sequence applied during the gate.
We consider $p$ pulse layers, applied at times $t_1,t_2, \dots, t_{p}$ corresponding to $k=p+1$ intervals of free time evolution between the pulses. 
Moreover, $t_0$ and $t_{p+1}=\tau$ are the initial and final gate times, respectively.
We denote the computational basis state during the $l$th time interval of the sequence as $s^{(l)}$ with components $s_i^{(l)}=\mathcal{S}_{il}s_i$. 
As discussed in Sec.~\ref{sec:Interaction_Engineering}, the decoupling pulse sequence is defined by the sign matrix $\mathcal{S}_n$, where we further use $\mathcal{S}_{i+n,l}=\mathcal{S}_{il}$ for $i = 1,\dots,n$ in the following.

Assuming a sequence with only a single pulse, the propagator after the full sequence reads $U(t_2,t_1) U_c^{(1)} U(t_1,t_0)$, corresponding to two intervals of free evolution separated by $U_c^{(1)}$ describing the $X$-pulses applied to the qubits. 
Accordingly, the fidelity after applying $p$ pulses reads
\begin{equation}
F = \frac{1}{2^{2N}} \sum_{\{s,r\}}e^{-\ci(\Theta_s - \Theta_r)}e^{\ci (\Delta_s - \Delta_r )}\prod_{m=1}^N\Big\langle D_m^\dagger(\beta^{(1)}_{m r}) \ldots D_m^\dagger(\beta^{(k)}_{m r})
D_m(\beta^{(k)}_{m s}) \ldots D_m(\beta^{(1)}_{m s}) \Big\rangle_{\mathrm{th}},
\end{equation}
where $\Theta_{s} = \sum_{l=1}^{k} \theta_{s^{(l)}}(t_{l},t_{l-1})$ represents the total phase accumulated over all intervals of free evolution, with each phase being $\theta_{s^{(l)}}(t_{l},t_{l-1})=\sum_{ij} s_i^{(l)} s_j^{(l)} \left[ (t_l-t_{l-1})J_{ij} + K_{ij}(t_l,t_{l-1}) \right] $.
Moreover, $\beta_{ms}^{(l)}= \sum_{j=1}^N \alpha_{jm}(t_l,t_{l-1})s_j^{(l)}$ denotes the displacement of mode $m$ associated with the computational basis state labeled by $s$ during the $l$th interval of free evolution. 
Through the interval-dependent states $s_i^{(l)}$, both the accumulated phases and the spin-dependent displacements now account for the sign changes induced by the applied $X$-pulse sequence.
Similarly to Eq.~\eqref{eqn:fidelity_free_evolution}, we evaluate the thermal expectation values and obtain the fidelity expression  underlying the results presented in the main text,
\begin{equation}\label{eq_fid_spectator_modes_exact}
F = \frac{1}{2^{2N}} \sum_{\{s,r\}}e^{-\ci(\Theta_s - \Theta_r)}e^{\ci (\Delta_s - \Delta_r )}\prod_{m=1}^N e^{\Gamma_{msr}}\exp\Bigg({ - \bigg| \sum_{l=1}^{k} \Big(\beta^{(l)}_{m s} - \beta^{(l)}_{m r}\Big) \bigg|^2(\bar{n}_m+1/2) }\Bigg),
\end{equation}
where $\Gamma_{msr}=\frac{1}{2} \sum_{j=1}^{2k} \sum_{i=j+1}^{2k} ( \gamma_j \gamma_i^* - \gamma_j^*\gamma_i) $ is defined in terms of $\gamma= \left(-\beta_{mr}^{(1)},  \ldots, -\beta_{mr}^{(k)},\beta_{ms}^{(k)}, \ldots, \beta_{ms}^{(1)} \right)$.
As before, the fidelity captures both coherent phase errors induced by spectator-mode coupling and fidelity loss due to residual qubit--motion entanglement, whose effect is enhanced by thermal mode occupation.
In addition, the interval-dependent phases and displacements account for the modifications induced by the applied decoupling sequence.

For sufficiently far-detuned spectator modes, the accumulated phase-space displacements remain small, ${\sum_{l=1}^k |\beta_{ms}^{(l)}| \ll 1}$.
Consequently, provided that the thermal occupations are not too large, both $\Gamma_{msr}$ and $\big| \sum_{l=1}^{k} (\beta^{(l)}_{m s} - \beta^{(l)}_{m r}) \big|^2 (\bar{n}_m + 1/2)$ are also small. 
Expanding Eq. \eqref{eq_fid_spectator_modes_exact}  then yields
\begin{equation}\label{eq_fidelity_spectator_modes_linearized}
    F \approx \frac{1}{2^{2N}} \sum_{\{s,r\}}e^{-\ci(\Theta_s - \Theta_r)}e^{\ci (\Delta_s - \Delta_r )} + \frac{1}{2^{2N}} \sum_{\{s,r\}}  \sum_m   \Gamma_{msr} - \frac{1}{2^{2N}} \sum_{\{s,r\}}  \sum_m  \bigg| \sum_{l=1}^{k} \Big(\beta^{(l)}_{m s} - \beta^{(l)}_{m r}\Big) \bigg|^2 (\bar{n}_m + 1/2).
\end{equation}
Here, we neglect mixed higher-order terms between the coherent phase errors $\Theta_s-\Delta_s$ and the displacement-induced corrections, assuming that both contributions are small.
As the decoupling sequence cancels the accumulated contribution of each unwanted coupling $J_{ij}$ at the end of the gate, only the transversal couplings $J_{i,i+n}$ contribute to the net evolution.
Moreover, the pulse sequence described by $\mathcal{S}_n$ is applied at equally spaced time intervals and acts symmetrically on both ion chains, such that opposing ion pairs are subject to the same pulse pattern. 
Consequently, the sign of the transversal interaction between opposing ions remains unchanged throughout the sequence, and $\Theta_s$ is independent of the sign pattern given by $\mathcal{S}_n$.
We therefore obtain $\Theta_s-\Delta_s \approx \sum_{i=1}^n s_{i} s_{i+n} \Delta\varphi_{i}$ with
\begin{equation}
    \Delta\varphi_i = \frac{\pi}{4} - 2 \tau \sum_{m=1}^{N} \frac{F_{i,m} F_{i+n,m}^* }{\delta_m} + 2k \sum_{m=3}^N \frac{F_{i,m} F_{i+n,m}^* }{\delta_m^2} \sin(\delta_m \tau/k ).
\end{equation}
The COM and STR modes do not contribute to the second sum since their phase-space trajectories are closed at the end of each decoupling interval. 
Note that $\Delta\varphi_i$ quantifies the deviation from the ideal entangling phase of ion pair $i,i+n$, including both the intrinsic phase error and the additional contribution induced by spectator-mode coupling.
The first term in Eq.~\eqref{eq_fidelity_spectator_modes_linearized} can therefore be written as $\frac{1}{2^{2N}} \sum_{\{s,r\}} e^{-\ci(\Theta_s - \Theta_r)} e^{\ci (\Delta_s - \Delta_r )} = \prod_{i=1}^{n} \cos^2(\Delta\varphi_i)$, while the second term vanishes since $\frac{1}{2^{2N}} \sum_{\{s,r\}} \sum_m \Gamma_{msr} = 0$.
With these considerations, we get 
\begin{equation}
F \approx \prod_{i=1}^{n} \cos^2(\Delta\varphi_i) - \sum_{m=1}^{2n} (2\bar{n}_m+1) \sum_{j=1}^{2n} |\tilde{\alpha}_{j,m}|^2,
\end{equation}
with the accumulated displacement of mode $m$ due to ion $j$,
\begin{equation}
    \tilde{\alpha}_{j,m} 
    = \sum_{l=1}^{k}\! \alpha_{j,m}^{(l)} \mathcal{S}_{jl} 
    =  \frac{g_{j,m}}{\delta_m} \sum_{l=1}^{k}  (e^{\ci \delta_m t_{l-1}} - e^{\ci \delta_m t_l}) \mathcal{S}_{jl},
\end{equation} 
where $\alpha_{j,m}^{(l)}$ denotes the corresponding displacement contribution accumulated during the $l$th decoupling interval.

\section{Refocusing of magnetic-field fluctuations}\label{app:magnetic_compensation}
In this appendix, we derive the conditions under which the applied decoupling sequences suppress dephasing caused by slowly varying magnetic-field fluctuations.
In the presence of magnetic-field fluctuations, the local dephasing Hamiltonian is given by
\begin{equation}
    H_\mathrm{B}(t)=-\frac{1}{2}\sum_i\delta\omega_i(t)Z_i,
\end{equation}
where $\delta\omega_i(t)=\delta\epsilon_{1,i}(t)-\delta\epsilon_{0,i}(t)$ denotes the fluctuation of the qubit transition frequency induced by magnetic-field fluctuations.
For a decoupling sequence consisting of $k$ equal time intervals of duration $\Delta t=\tau/k$, the sign of the dephasing term during interval $l$ is determined by $\mathcal{S}_{il}$, where $\mathcal{S}$ is the decoupling matrix (cf. Sec.~\ref{sec:Interaction_Engineering}).
The phase accumulated by ion $i$ over the full gate duration is therefore
\begin{equation}
    \varphi_i=-\sum_{l=1}^{k}\mathcal{S}_{il}\int_{(l-1)\Delta t}^{l\Delta t}\delta\omega_i(t)\,\mathrm{d}t.
\end{equation}
For a frequency shift varying linearly during the gate, $\delta\omega_i(t)=\delta\omega_{i,0}+\delta\omega_{i,1}t$, the constant and linear contributions to the accumulated phase vanish independently if
\begin{equation}
    \sum_{l=1}^{k}\mathcal{S}_{il}=0,
    \qquad
    \sum_{l=1}^{k}(2l-1)\mathcal{S}_{il}=0,
\end{equation}
respectively.
These conditions correspond to vanishing zeroth and first temporal moments of the sign sequence.
The first condition is satisfied by all Walsh functions except $W_0$, since all nonconstant Walsh functions contain equal numbers of positive and negative entries.
In contrast, cancellation of the linear contribution depends on the temporal ordering of the signs and therefore on the specific Walsh function.

\end{document}